%% file: article_arxiv.tex
\documentclass[final,3p,12pt,times]{elsarticle}

\biboptions{authoryear}

\usepackage{graphicx}
\usepackage{enumitem}
\usepackage{natbib}
\usepackage{url} 
\usepackage{amssymb,amsfonts}
\usepackage{amsthm}
\usepackage{amsmath}
\usepackage{booktabs}
\usepackage{verbatim}
\usepackage{bm,bbm}

\usepackage{multibib}
\newcites{apndx}{References in Online Appendix}
\newcommand{\blind}{1}

\usepackage{xcolor}
\usepackage{subcaption}
\usepackage{setspace}
\newtheorem{theorem}{Theorem}
\newtheorem{proposition}[theorem]{Proposition}%

\journal{t.b.d.}

\input{notation.tex}

\def\mytitle{Spectral Dynamics and Regularization for High-Dimensional Copulas}

\begin{document}

\begin{frontmatter}
\title{\mytitle}

\author[UvT]{Koos B. Gubbels}
\ead{k.b.gubbels@tilburguniversity.edu}
          
\author[VU]{Andre Lucas}
\ead{a.lucas@vu.nl}

\affiliation[UvT]{organization={Department of Econometrics and Operations Research, Tilburg University},
            city={Tilburg},
            country={The Netherlands}}

\affiliation[VU]{
            organization={Department of Econometrics and Data Science and Tinbergen Institute,  Vrije Universiteit Amsterdam},
            city={Amsterdam},
            country={The Netherlands}}


\begin{abstract}
We introduce a novel model for time-varying, asymmetric, tail-dependent copulas in high dimensions that incorporates both spectral dynamics and regularization.
The dynamics of the dependence matrix' eigenvalues are modeled in a score-driven way, while biases in the unconditional eigenvalue spectrum are resolved by non-linear shrinkage.
The dynamic parameterization of the copula dependence matrix ensures that it satisfies the appropriate restrictions at all times and for any dimension.
The model is parsimonious, computationally efficient, easily scalable to high dimensions, and performs well for both simulated and empirical data.
In an empirical application to financial market dynamics using 100 stocks from 10 different countries and 10 different industry sectors, we find that our copula model captures both geographic and industry related co-movements and outperforms recent computationally more intensive clustering-based factor copula alternatives.
Both the spectral dynamics and the regularization contribute to the new model's performance.
During periods of market stress, we find that the spectral dynamics reveal strong increases in international stock market dependence, which causes reductions in diversification potential and increases in systemic risk.\\[1ex]\end{abstract}


\end{frontmatter}


\section{Introduction}
\label{sec:intro}

High-dimensional models for dependence play an important role in quantitative risk management and finance; for an overview, see, e.g., \citet{McNeil2005}. 
In such high-dimensional settings, standard multivariate densities are typically too tightly parameterized to describe the data well.
The standard solution to this problem, which is also adopted in this article, is to use the more flexible copula perspective.
Here, one splits the modeling process into two steps: a first stage where one builds univariate models for the marginal properties of each of the observed time series, and a second stage where one formulates a copula to capture the multivariate dependence structure between the different series \citep[see for instance][]{Joe2014}.

To capture financial market data well, there are three stylized facts that any good copula model should pick up, namely (i) time-variation of the dependence structure, (ii) asymmetry, and (iii) tail-dependence.
In addition, a good high-dimensional copula model should be able to deal with (iv) the increase in the number of parameters as the dimensionality grows, and (v) the biases arising in a high-dimensional context for typical dependence measures like covariance matrices or copula dependence matrices \citep[see, for instance,][]{LedoitWolf2004,LedoitWolf2012}.
Earlier literature typically deals with some, but not all of these challenges simultaneously, and a unified model dealing with all these challenges at the same time is currently lacking. 
For instance, \citet{lucas2017modeling} use a dynamic, skewed and fat-tailed copula model for the time-varying dependence between 73 sovereigns and banks, but adopt a highly restrictive block-equicorrelation structure as in \citet{engle2012dynamic} to keep the number of parameters manageable in high dimensions.
\cite{Engle2019} on the other hand consider a vast-dimensional setting and use non-linear shrinkage to overcome the biases in the estimation of the long-term mean of the dynamic correlation matrix in the DCC specification of \citet{Engle2002}.
They do not take a copula perspective, however, and do not account for possible skewness in the dependence structure. 
Moreover, if $\dimN$ denotes the number of assets, their dynamics are imposed on the entire $\dimN\times\dimN$ correlation matrix via the DCC specification, rather than on the dynamics of the much lower dimensional spectrum itself.

A different stream of literature use a factor-based copula approach with exogenous \citep{Creal2015,Oh2017,Opschoor2020} or endogenous \citep[via clustering;][]{Oh2023} assignment of cross-sectional units to groups with similar factor exposures.
These factor copulas are typically dynamic and fat-tailed (and sometimes skewed), whereas the dimensionality challenges are tackled via the allocation of assets to groups and the pooling of factor loadings.
Such group structures may, however, not be exact or may not even exist in particular applications, in which case a group-based factor copula model approach can become suboptimal or even biased.
Moreover, endogenously deciding on the number of clusters and factors in the context of a non-linear dynamic model can be quite challenging and become computationally prohibitively expensive in high dimensions.

In this article, we therefore propose a new high-dimensional copula model that addresses all the above challenges in one unified framework.
Our copula model includes skewness and tail dependence by starting from the generalized hyperbolic skewed $t$ copula as in, for instance, \citet{lucas2017modeling}.
We address the challenge of high-dimensionality by focusing on the spectral decomposition of the dependence matrix, modeling the dominant eigenvalues as dynamic, while keeping the remaining eigenvalues static and debiasing the eigenvalue spectrum using non-linear shrinkage techniques developed by \cite{Ledoit2022a, Ledoit2022b}.
\citet{hetland2023dynamic} study eigenvalue dynamics of a covariance matrix in a low-dimensional context and find that the score-driven dynamics of \cite{Creal2013} and \citet{Harvey2013} work well for eigenvalues.
We extend their framework (i) to a copula context with skewness and fat-tailedness, and (ii) to a high-dimensional context by including the regularization techniques required for debiasing the spectrum \citep{Ledoit2004,Ledoit2022a,Ledoit2022b}. 
In contrast to \cite{hetland2023dynamic}, we explicitly distinguish between the marginal and the copula time series. 
Therefore, our dynamic eigenvalues solely reflect the dependence structure, allowing us to show that market movements with increased dependence are followed by subsequent periods with high dependence. 

We show in a simulation study that our copula model can recover cluster-based dependence structures with limited performance loss compared to cluster factor copulas of \citet{Oh2023} if the latter form the true data generating process (dgp).
However, when the dependence structure increasingly deviates from an exact group structure, our high-dimensional copulas pick up the dependence structure more accurately than the cluster-based factor copula alternative.
In an application to the dynamic dependence structure of global financial markets using 100 stocks from 10 different countries and 10 different industry sectors, we corroborate these results. 
So far, high-dimensional copula studies have mainly focused on stocks from a single country, such as the US \citep{Oh2023}. 
The large number of possible country and sector combinations in our current application complicates the detection of a proper factor structure using standard clustering approaches.
We find that high-dimensional copulas with regularized spectral dynamics perform well under these challenging circumstances.
For the empirical data considered in the application, both the dynamics of the spectrum and the regularization of the spectrum contribute to the new model's performance.
Moreover, the spectral copula dynamics reveal that in times of financial distress the stock market dependence structure stretches along the first spectral dimension, reducing diversification possibilities and increasing systemic risk.

The remainder of the paper is structured as follows. 
In Section \ref{sec:Def} we introduce the modeling framework and discuss regularization of the dynamic spectra and estimation of the static parameters.
Section \ref{sec:Sim} presents simulation evidence.
Section \ref{sec:Emp} applies the new model to global financial markets. 
Section \ref{sec:Concl} concludes.  

\section{The model}
\label{sec:Def}


\subsection{Generalized hyperbolic skewed \textit{t} copula}
\label{subsec:GH}

Consider a vector-valued time series $\vyt = (\dy_{1,\dimt},\ldots,\dy_{\dimN,\dimt})\trans \in \SR^{\dimN\times 1}$ observed for $\dimt=1,\ldots,\dimT$.
For each $\dyit$, assume the availability of a conditional marginal model in the form of a cumulative distribution function (cdf) $\cdfyi(\cfdot|\calFtm)$ for a common information set $\calFt = \{\vy_s\}_{s\le\dimt}$; see \citet{patton2006modelling}.
We use these marginal models to construct conditional probability integral transforms (PITs) $\vPITt = (\PIT_{1,\dimt},\ldots,\PIT_{\dimN,\dimt})\trans \in \SR^{\dimN \times 1}$ with $\dPITit = \cdfyi(\dyit\mid\calFtm)$, whose multivariate distribution is the main focus in the remainder of our analysis.
We assume the following dynamic conditional copula specification for the PITs:
\begin{align}
    \label{eq:GH copula}
    \vPITt &\sim
    \copulac(\vPITt\mid \calFtm, \vthetact)
    =
    \copulac(\vPITt; \vthetact)
    =
    \frac{
        \GHpdf\bra{ 
            \GHcdf_1^{-1}(\PIT_{1,\dimt}; \vthetact), 
            \ldots, 
            \GHcdf_{\dimN}^{-1}(\PIT_{\dimN,\dimt}; \vthetact) 
        }
    }{
        \GHpdf_1\bra{ \GHcdf_1^{-1}(\PIT_{1,\dimt}; \vthetact) }
        \cdots
        \GHpdf_{\dimN}\bra{ \GHcdf_{\dimN}^{-1}(\PIT_{\dimN,\dimt}; \vthetact) }
    }
    ,
\end{align}
where the dynamic copula parameter $\vthetact$ summarizes all the dependence information from the information set $\calFtm$ as needed for the copula, and where $\GHpdf(\cfdot)$ is an appropriate multivariate density with marginal pdfs $\GHpdf\subi(\cfdot;\cfdot)$ and cdfs $\GHcdf\subi(\cfdot;\cfdot)$ for $\dimi=1,\ldots,\dimN$.
As an appropriate choice for $\GHpdf(\cfdot;\cfdot)$, we consider the multivariate generalized hyperbolic skewed $t$ copula with degrees-of-freedom parameter $\GHnu$, skewness parameter $\GHvgamma$, and scale matrix $\GHmRt = \GHmR(\vthetact)$, as generated by
\begin{align}
    \label{eq:GH pdf}
    \GHpdf\bra{\vyst; \vthetact}
    &=
    \frac{
        2^{-\GHnu/2+1}\GHnu^{\GHnu/2}\GHalphat^{(\dimN+\GHnu)/2}
    }{
        (2\pi)^{d/2}\Gamma(\GHnu/2)|\GHmRt|^{1/2}
    } 
    \frac{
        e^{\vyst{}\trans\GHvbetat}
        K_{(\GHnu+\dimN)/2} \bra{
            \GHalphat \sqrt{\GHnu+\vyst{}\trans\GHmRt^{-1}\vyst}
        }
    }{
        \bra{\GHnu+\vyst{}\trans\GHmRt^{-1}\vyst}^{(\dimN+\GHnu)/4}
    }
    ,
\end{align}
where $\vyst = (\dys_{1,\dimt},\ldots,\dys_{\dimN,\dimt})\trans$, with $\dysit = \GHcdf_{\dimi}^{-1}\bra{\dPITit; \vthetact}=\GHcdf_{\dimi}^{-1}\bra{\dPITit; \GHnu,\GHgammai}$ for $\dimi=1,\ldots,\dimN$, and where $\GHvbetat=\GHmRt^{-1}\GHvgamma$ and $\GHalphat = (\GHvgamma\trans\GHmRt^{-1}\GHvgamma)^{1/2}$; see also \citet{Demarta2005}.
As the marginal scales of the copula are not identified, we restrict the diagonal of $\GHmRt$ to be equal to one, such that $\GHmRt$ has a correlation matrix form.

The multivariate generalized hyperbolic skewed $t$ distribution in \eqref{eq:GH pdf} can be constructed as a mean-variance mixture
\begin{align}
	\label{eq:GH mv mixture}
	\vyst &=
	\dvt\,\GHvgamma + \sqrt{\dvt}\, \GHmRt^{1/2}\vzt,
\end{align}
where $\vzt \sim \Nn(\vzeroN, \unitmatN)$ is a multivariate standard normally distributed random variable, and $\dvt \sim {\rm IG}(\GHnu/2,\GHnu/2)$ is Inverse Gamma distributed and independent of $\vzt$.
Here $\GHmRt^{1/2}$ denotes the symmetric root of $\GHmRt$.
The above distribution has been a popular choice to study dynamic dependence structures in financial markets \citep{Lucas2014,Opschoor2020, Oh2023}.
The distribution is both flexible and analytically tractable and accommodates skewness as well as fat tails. 
It is immediately clear from \eqref{eq:GH mv mixture} that the marginal distribution of $\dysit$ is univariate skewed $t$ with shape parameter $\GHnu$, skewness parameter $\GHgamma_\dimi$, and scale parameter 1.
The marginal pdfs $\GHpdf_i(\cfdot;\cfdot)$ in \eqref{eq:GH copula} are thus known analytically.

\subsection{Spectral dynamics}
\label{subsec:dynamics and shrinkage}

The core challenge for a high-dimensional time-varying copula model is the curse of dimensionality and the explosion of the number of free parameters in $\GHmRt$.
Different solutions have been proposed.
\citet{Oh2018} and \citet{Opschoor2020} study factor copulas where $\GHmRt$ is decomposed into a low-rank matrix plus a diagonal matrix, both of which can vary over time.
Further parsimony can be imposed by pooling the dynamic parameters across pre-specified groups, for example, industries, or by choosing the groups in a data-driven way, e.g., using clustering techniques as in \citet{Oh2023}.
Clustering in the context of a non-linear model can quickly become time-consuming if either the sample size or the number of assets becomes large.
Moreover, clustering techniques may face challenges if the cluster structure is only approximate; see Section~\ref{sec:Sim} for the effects of such deviations in a controlled setting.

To overcome these issues, we do not impose a group structure, but instead use a normalized spectral decomposition of the copula dependence matrix $\GHmRt$ by imposing
\begin{align}
    \label{eq:R parameterization}
    \GHmRt 
    &=
    \diag\bra{
        \mW \mLambdat \mW\trans
    }^{-1/2}\ 
    \mW \mLambdat \mW\trans
    \diag\bra{
        \mW \mLambdat \mW\trans
    }^{-1/2}
    ,
\end{align}
where $\mW$ is orthogonal and $\mLambdat$ is time-varying and diagonal with strictly positive entries.
Note that the parameterization of $\GHmRt$ automatically ensures that $\GHmRt$ has the format of a correlation matrix as long as the diagonal elements of $\mLambdat$ are strictly positive.
The latter can be ensured by using an exponential link function for the diagonal elements of $\mLambdat$.
Alternatives for parameterizing dynamic correlation matrices are, for example, the hypersphere parameterization of \citet{Jaeckel2000} as used in \citet{Creal2011} and \citet{buccheri2021score}, or the log-correlation matrix parameterization of \citet{ArchakovHansen2021} as used in \citet{HafnerWang2023}.
By concentrating on the eigenvalues, the number of free dynamic parameters is considerably less than in these alternative parameterizations, which helps for the model's tractability in high-dimensional settings.
Moreover, the spectral parameterization of the correlation matrix in \eqref{eq:R parameterization} still allows for explicit derivative expressions with respect to all nonzero elements in $\mLambdat$.

We describe the dynamics of $\mLambdat$ using the score-driven approach of \citet{Creal2013} and \citet{Harvey2013},
\begin{align}
    \label{eq:first score eq}
    \vftp &= \vomega + \mB\,\vft + \mA\,\vnablat, 
\end{align}
where $\vnablat = \partial\GHpdf(\vyst;\vthetact)/\partial\vft$ is the score of the copula density, $\vft = \log\vlambdat$, and where we use unit scaling as defined by \citet{Creal2013}.
The relevant score equations for the new parameterization from \eqref{eq:R parameterization} in the copula setting are given by the following proposition, the proof of which can be found in the appendix. 

\begin{proposition}[\textbf{score equations for spectral dynamics}]
    \label{prop:score dynamics equations}
    Let $\copulac\bra{\vPITt; \vthetact}=\copulac\bra{\vPITt; \GHmRt, \GHnu, \GHvgamma}$ be the skewed $t$ copula density from \eqref{eq:GH copula} and \eqref{eq:GH pdf}, let $\dfit = \log\dlambdait$, $\dlambdait = \Lambda_{i,i,t}$ and $\GHalphatildet = \GHalphat\sqrt{\GHnu + \vyst{}\trans\GHmRt^{-1}\vyst}$ with $\GHalphat = \sqrt{\GHvgamma\trans\GHmRt^{-1}\GHvgamma}$. 
    Then, using $\vnablat = (\dnabla_{1,\dimt},\ldots,\dnabla_{\dimN,\dimt})\trans = \partial \log \copulac\bra{\vPITt; \vthetact}/\partial \vft$, we have that the unit-scaled score dynamics are given by Eq.~\eqref{eq:first score eq}, with
    \begin{align}
        \label{eq:GH score expression}
        \dnablait &=
        \frac{\partial \log \copulac(\vPITt; \GHmRt,\GHnu,\GHvgamma)}{\partial \dfit}
        = 
        -\tfrac12\ 
        \frac{\partial \log|\GHmRt|}{\partial \dfit} 
        -\tfrac12\ 
        \vyst{}\trans
        \frac{\partial\GHmRt^{-1}}{\partial\dfit}
        \bra{\dwt\ \vyst - 2\GHvgamma}
        \\
        \nonumber
        &\qquad\qquad
        +
        \tfrac12\,
        \GHalphatildet\cdot
        k'_{(\GHnu+\dimN)/2}\bra{\GHalphatildet} \times
        \Bigg(
            \frac{\GHvgamma\trans
            \bra{\partial\GHmRt^{-1}/\partial\dfit}\GHvgamma}{\GHvgamma\trans \GHmRt^{-1}\GHvgamma}
            +
            \frac{\vyst{}\trans \bra{\partial\GHmRt^{-1}/\partial\dfit}\vyst}{\nu+\vyst{}\trans \GHmRt^{-1}\vyst}
        \Bigg)
        ,
    \end{align}
    where $\dwt = (\dimN + \GHnu)/(\GHnu + \vyst{}\trans\GHmRt^{-1}\vyst)$, and
    \begin{align}
        \nonumber
        &\GHmRt = 
        \diag(\mSigmat)^{-1/2}\ \mW \mLambdat \mW\trans\ \diag(\mSigmat)^{-1/2},
        \qquad
        \mSigmat = \mW \mLambdat \mW\trans,
        \qquad
        \\
        \nonumber
        &\frac{\partial \log|\GHmRt|}{\partial \dfit} 
        = 1-\dlambdait \sum_{\dimj=1}^{\dimN}
        \frac{\dW^2_{\dimj,\dimi}}{\Sigma_{j,j,t}},
        \qquad
        \mSigmadotit = \tfrac12\dlambdait
        \diag\bra{
            \frac{\dW_{1,\dimi}^2}{\dSigma_{1,1,\dimt}}
            , \ldots,
            \frac{\dW_{\dimN,\dimi}^2}{\dSigma_{\dimN,\dimN,\dimt}}
        },
        \\
        \nonumber
        &\frac{\partial\GHmRt^{-1}}{\partial\dfit}
        = 
        -\diag(\mSigmat)^{1/2}\, \frac{\vw\subi\vw\subi\trans}{\dlambdait}\, \diag(\mSigmat)^{1/2} 
        +
        \mSigmadotit\,\GHmRt^{-1} +
        \GHmRt^{-1}\,\mSigmadotit,
    \end{align}
    and $k_{\GHnu}'(x) = \partial \log \bra{x^{\nu}\cdot K_{\GHnu}(x)}/\partial x$ for given $\GHnu > 0$ and $x\in\SR^+$, where $\vw\subi$ denotes the $\dimi$th column of $\mW$ and $\dW_{\dimj,\dimi}$ its $(\dimj,\dimi)$th element, and $\dSigma_{\dimj,\dimj,\dimt}$ denotes the $(\dimj,\dimj)$th element of $\mSigmat$.
\end{proposition}

The eigenvalue dynamics in Proposition~\ref{prop:score dynamics equations} are substantially different from the Gaussian and Student's $t$ based eigenvalue dynamics in  \citet{hetland2023dynamic} in at least three respects: the dynamics in Proposition~\ref{prop:score dynamics equations} relate to (i) a high-dimensional copula setting, (ii) a correlation matrix rather than a covariance matrix parameterization, and (iii) a more general class of distributions.
Still, the dynamics of $\vlambdat$ given in Proposition~\ref{prop:score dynamics equations} have an intuitive form.
To see this, we introduce a scaled and rotated version of $\vyst$, namely $\vystildet = \mW\trans\diag(\mSigmat)^{1/2}\vyst$.
We also define the corresponding rescaled and rotated version of $\GHvgamma$, namely $\GHvgammatildet = \mW\trans\diag(\mSigmat)^{1/2}\GHvgamma$.
We can now rewrite \eqref{eq:GH score expression} as
\begin{align}
    \dnablait
    =&
    \tfrac12\,\bra{
        \dwt
        \frac{\dystildeit{}^2}{\dlambdait}
        - 1
    }
    -
    \frac{\GHgammatildeit\dystildeit}{\dlambdait}
    -\tfrac12\,
    \GHalphatildet\cdot
    k'_{(\GHnu+\dimN)/2}\bra{\GHalphatildet} \cdot
    \bra{
        \frac{\GHgammatildeit^2/\dlambdait}{\GHvgammatildet\trans\mLambdat^{-1}\GHvgammatildet}
        +
        \frac{\dystildeit{}^2/\dlambdait}{\nu+\vystildet{}\trans\mLambdat^{-1}\vystildet}
    }
    \nonumber \\      
    &-\,\bra{
        \dwt\,
        \vystildet{}\trans\mLambdat^{-1}\vysbarit
        - 
        \trace(\mSigmadotit)
    }
    +
    \bra{
        \GHvgammatildet{}\trans\mLambdat^{-1}\vysbarit
        +
        \GHvgammabarit\trans\mLambdat^{-1}\vystildet
    } \nonumber \\ 
    &+
    \GHalphatildet\cdot
    k'_{(\GHnu+\dimN)/2}\bra{\GHalphatildet} \cdot
    \bra{
        \tfrac{\GHvgammatildet\trans\mLambdat^{-1}\GHvgammabarit}{\GHvgammatildet\trans\mLambdat^{-1}\GHvgammatildet}
        +
        \tfrac{\vystildet{}\trans\mLambdat^{-1}\vysbarit}{\nu+\vystildet{}\trans\mLambdat^{-1}\vystildet}
    }\label{eq:GH score expression rewrite}
    ,
\end{align}
where $\vysbarit$ and $\GHvgammabarit$ are similar rescaled and rotated versions of $\vyst$ and $\GHvgamma$, respectively, as defined in the proof of \eqref{eq:GH score expression rewrite} in \ref{app:proofs}.

The score in \eqref{eq:GH score expression rewrite} consists of two main parts: terms 1 to 3, and terms 4 to 6.
The first term holds the familiar weighted EGARCH-like volatility dynamics $\dwt\/\dystildeit{}^2/\dlambdait - 1$ that is familiar from the literature on score-driven models for a time-varying log-variance \citep[see][]{Creal2013,Harvey2013}.
This is intuitive, as the $\dimi$th eigenvalue is the variance of the the $\dimi$th spectral projection $\dystildeit$.
The weighting factor $\dwt$ causes extreme observations to have less impact on the eigenvalue dynamics for finite $\GHnu$, and it collapses to unity if $\nu\to\infty$.
The second and third term of \eqref{eq:GH score expression rewrite} are due to the skewness of the copula specification.
They vanish if $\GHgammatildeit$ equals zero.
If $\GHgammatildeit > 0$, then a large positive $\dystildeit$ has a smaller impact on the next eigenvalue $\dlambda_{\dimi,\dimt+1}$ than a large negative $\dystildeit$.
This is because large positive outcomes are more likely under positive skewness and are, therefore, not attributed to volatility increases along the corresponding spectral dimension.
For negative skewness, the opposite holds. 

The second main part of the score in \eqref{eq:GH score expression rewrite}, terms 4--6, stem from our parameterization of $\GHmRt$ as a correlation matrix.
These three terms mimic the first three terms, but with an opposite sign.
We first see a weighted EGARCH type term $(\dwt\vystildet{}\trans\mLambda^{-1}\vysbarit - \trace(\mSigmadotit)$, followed by a term related to the skewness, and a term related to the combination of skewness and kurtosis.
These additional, more complex terms adjust the dynamics of the eigenvalues to account for the fact that $\GHmRt$ has unit diagonal elements by construction.
They therefore involve all the spectral projections in $\vystildet$ simultaneously, combined with a related spectral projection $\vysbarit$ that includes an additional scaling by $\mSigmadotit$.
Despite its more complex expression, the score is still easy to compute and available in analytical form and decomposable into different terms that are attributable to the features of the skewed $t$ copula.

Due to the explicit normalization of $\GHmRt$ in Eq.~\eqref{eq:R parameterization} and Proposition~\ref{prop:score dynamics equations}, $\GHmRt$ is a proper correlation matrix for any update of $\vft$. 
The number of time-varying parameters in $\GHmRt$, however, is considerably less than in a full dynamic hypersphere or log-correlation matrix parameterization as in \citet{Creal2011}, \citet{buccheri2021score}, \citet{ArchakovHansen2021}, or \citet{HafnerWang2023}.
Also note that not all parameters in the parameterization of $\GHmRt$ in \eqref{eq:R parameterization} can be identified simultaneously.
For instance, both $\vlambdat$ and $k\cdot\vlambdat$ for some constant $k>0$ give the same copula dependence matrix $\GHmRt$ in \eqref{eq:R parameterization}.
We solve this by restricting the elements of $\vomega$ in the recurrence relation for $\vftp$ later on using a targeting approach.

The spectral copula approach of Proposition~\ref{prop:score dynamics equations} comes with several advantages.
First, the approach is purely data-driven, as opposed to using a pre-defined group classification based on, for instance, industries as in \citet{Oh2018} and \citet{Opschoor2020}.
Second, the spectral decomposition is computationally much less demanding compared to a full clustering-based approach as in, for example, \citet{Oh2023}.
Third, the approach imposes no restrictions on the dependence matrix.
Fourth, the spectral approach facilitates an exploratory phase of the modeling process.
For instance, using an initial guess of $\GHnu$ and $\GHvgamma$ (such as the standard normal $\GHvgamma = \vzero$ and $\GHnu\to\infty$), the inverse cdf transforms $\vyst$ of the PITs $\vPITt$ immediately lead to an initial estimate of $\hat{\mW}$ based on the unconditional correlation matrix of $\vyst$. 
This, in turn, allows us to construct preliminary estimates of the spectral projections of the data in $\dystildeit$, which one can inspect visually to get an impression of the extent of time-variation in the volatility of $\dystildeit$, i.e., in $\dlambdait$.
We can use such information to impose further parsimony on the model, e.g., by restricting some of the spectral dimensions to have constant rather than time-varying volatility.
We come back to this in Section \ref{subsec:data}.

\subsection{Non-linear shrinkage and model selection}
\label{subsec:nonlinear shrinkage}

In high-dimensions it becomes challenging to estimate the copula correlation matrix.
It is known that for large dimension $\dimN$ the sample eigenvalues $\{\vlambdahatt\}_{\dimt=1}^{\dimN}$ become a biased estimator of the true spectrum $\{\vlambdat\}_{\dimt=1}^{\dimN}$ when $\dimN,\dimT\to\infty$ and $\dimN/\dimT$ converges to some positive, nonzero constant \citep{marcenkopastur1967,Johnstone2001Spiked,LedoitWolf2012,Ledoit2022a,Ledoit2022b}.
\citet{Engle2019} use shrinkage techniques to correct these biases by adjusting the (targeted) high-dimensional intercept in their DCC transition equation.
They show that such a shrinkage procedure for the $\dimN\times\dimN$ volatility intercept provides better results than a targeting procedure without shrinkage.
We follow a similar procedure, using the de-biased spectrum based on the more recent quadratic shrinkage techniques of \citet{Ledoit2022b} as our target.
If $\vlambdahatt$ denotes the sample spectrum at time $t$, the quadratic shrinkage formula $ f_{\rm QS}$ is given by
\begin{align}
    \dmu\subit^{-1} 
    &= 
    f_{\rm QS}(\vlambdahatt^{-1})
    =
    \bra{1-\dq}^2\dlambdahat\subit^{-1} 
    + 
    2\dq\bra{1-\dq} \dlambdahat\subit^{-1}\frac1{\dimN}
    \sum_{\dimj=1}^{\dimN}
        \dlambdahat\subjt^{-1} 
        \frac{
            \dlambdahat\subjt^{-1} - \dlambdahat\subit^{-1}
        }{
            (\dlambdahat\subjt^{-1} - \dlambdahat\subit^{-1})^2
            +
            \dh^2\dlambdahat\subjt^{-2}
        }
    \nonumber
    \\
    \label{eq:Shrink}
    & \qquad
    +
    \dq^2\dlambdahat\subit^{-1}
    \bra{
        \sbra{
            \frac1{\dimN} \sum_{\dimj=1}^{\dimN}
            \dlambdahat\subjt^{-1} 
            \frac{
                \dlambdahat\subjt^{-1} - \dlambdahat\subit^{-1}
            }{
                (\dlambdahat\subjt^{-1} - \dlambdahat\subit^{-1})^2
                +
                \dh^2\dlambdahat\subjt^{-2}
            }
        }^2
        +
        \sbra{
            \frac1{\dimN} \sum_{\dimj=1}^{\dimN}
            \dlambda\subjt^{-1} 
            \frac{
                \dh\,\dlambdahat\subjt^{-1}
            }{
                (\dlambdahat\subjt^{-1} - \dlambdahat\subit^{-1})^2
                +
                \dh^2\dlambdahat\subjt^{-2}
            }
        }^2
    }
    ,
\end{align}
where $\vmu\subt = \big(\dmu_{1,\dimt},\ldots,\dmu_{\dimN,\dimt}\big)\trans$ denotes the de-biased spectrum, $\dq=\dimN/\dimT$, and $\dh$ is a smoothness parameter. 
\cite{Ledoit2022b} show that this type of shrinkage mapping is asymptotically optimal under various loss metrics, such as Frobenius loss.
They advise to take $\dh = \min(q^2, q^{-2})^{0.35}\cdot \dimN^{-0.35}$ and demonstrate that their non-linear shrinkage estimator has excellent in-sample and out-of-sample performance in a wide range of settings.

In combination with the above shrinkage approach to overcome eigenvalue biases in high dimensions, we also impose parsimony on the model by limiting the number of dynamic eigenvalues.
We do so using standard model-selection criteria.
In applications to stock return data as in Section~\ref{sec:Emp}, one typically finds that the first few eigenvalues are much larger than the remaining ones.
The notion of a few dominant eigenvalues that drive the dependence structure is conceptually in line with the perspective of a `spiked' eigenvalue dependence model \citep[see, e.g.,][]{Johnstone2001Spiked,FanLiaoMincheva2013POET,donoho2018optimal} and corresponds closely to the typical covariance structure present in financial markets \citep[see, for instance,][and many more]{FamaFrench1993,FamaFrench1998International}.
Capturing the dynamics of the largest eigenvalues is therefore most important and contributes most to the model's fit.
For the lowest eigenvalues, it is more important to prevent them from becoming too close to zero due to the high-dimensional biases, which would result in poor out-of-sample performance.
To select the number of dynamic eigenvalues, we therefore implement the following model selection strategy.
Starting from a model with only static eigenvalues, we increase the number of dynamic eigenvalues one-by-one, starting from the largest eigenvalue.
We then compute the change in BIC relative to the static model by accounting for the change in the in-sample log-likelihood and the change in the number of parameters. 
We select the dynamic model with the lowest BIC.
The procedure typically results in a limited number of dynamic eigenvalues and a large number of static ones.
Combined with the high-dimensional shrinkage procedure for targeting the intercepts, this results in a considerable reduction of the parameter space.
We show later using simulated and empirical data that this approach efficiently captures the salient features of the data, including their dynamics.

\subsection{Parameter estimation}
\label{subsec:parameter estimation}

We split the estimation procedure in two standard steps.
We first estimate the marginal behavior of each original univariate series $r_{i,t}$ by an AR(1)-GARCH(1,1) model using standard  quasi maximum likelihood (QMLE).
From the devolatilized residuals $\dyit=\epsilon_{i,t}/\sigma_{i,t}$, we use the non-parametric rank transformation $\dPITit = \text{rank}(\dyit)/(T+1/2)$ to obtain the PITs.
This makes the construction of the PITs more robust to any potential mis-specification of the marginal distributions, in line with the use of QMLE to estimate the AR(1)-GARCH(1,1) marginal models.

Next, we estimate the dynamic dependence structure by maximizing the copula density \eqref{eq:GH copula}.
For a given tail shape parameter $\GHnu$ and skewness parameter $\GHvgamma$, we can compute the inverse marginal cdf projections $\vyst = \vyst(\GHnu,\GHvgamma) = \GHcdf^{-1}(\dPITit;\GHvgamma,\GHnu)$.
Let $\Covys = \Ee[\vyst \vyst{}\trans]$ and $\GHmR = \Ee[\GHmRt]$, then for the skewed $t$ distribution we have that
\begin{align}
    \label{eq:target omega}
	\Covys &= 
	\frac{\GHnu}{\GHnu-2} \GHmR  + 
	\frac{2\GHnu^2}{(\GHnu-2)^2(\GHnu-4)}
	\GHvgamma\GHvgamma\trans
    \quad \Leftrightarrow \quad
    \GHmR =
	\frac{\GHnu-2}{\GHnu} \Covys -
	\frac{2\GHnu}{(\GHnu-2)(\GHnu-4)}
	\GHvgamma\GHvgamma\trans
	.
\end{align}
Replacing $\Covys$ in \eqref{eq:target omega} by the sample estimator $\Covyshat = \dimT^{-1}\sum_{\dimt=1}^{\dimT} \vyst\vyst{}\trans$, we then define the estimators
\begin{equation}
    \label{eq:target omega2}
    \begin{split}
    \hat\mSigma &=
    \hat\mSigma(\GHvgamma,\GHnu) =
    \hat\mW(\GHvgamma,\GHnu)\ \hat\mLambda(\GHvgamma,\GHnu)\ \hat\mW(\GHvgamma,\GHnu)\trans
    =
	\frac{\GHnu-2}{\GHnu} \Covyshat -
	\frac{2\GHnu}{(\GHnu-2)(\GHnu-4)}
	\GHvgamma\GHvgamma\trans
	,
    \\
    \hat\GHmR &= \diag(\hat\mSigma)^{-1/2}\ \hat\mSigma\ \diag(\hat\mSigma)^{-1/2}.
    \end{split}
\end{equation}
The intercepts of the score-driven transition Eq.~(\ref{eq:first score eq}) are targeted as explained in Section~\ref{subsec:nonlinear shrinkage} using the quadratic shrinkage procedure of \citet{Ledoit2022b} from Eq.~(\ref{eq:Shrink}), yielding the final (shrunken) estimators
\begin{align}
    \label{eq:target omega3}
        \mSigmashrink &=
    \mSigmashrink(\GHvgamma,\GHnu) =
    \hat\mW(\GHvgamma,\GHnu)\ \mLambdashrink(\GHvgamma,\GHnu)\ \hat\mW(\GHvgamma,\GHnu)\trans
    ,
    &
    \GHmRshrink &= 
    \diag(\mSigmashrink)^{-1/2}\ \mSigmashrink\ \diag(\mSigmashrink)^{-1/2}
    ,
\end{align}
where $\mLambdashrink = \mLambdashrink(\GHvgamma,\GHnu)$ holds the quadratically shrunken spectrum from Eq.~(\ref{eq:Shrink}).
Note that all these quantities are still functions of $\GHvgamma$ and $\GHnu$.
Finally, we compute the copula log-likelihood function as
\begin{align}
    \label{eq:loglik iteration k}
    \LL(\vstaticpar) &=
    \sum_{\dimt=1}^\dimT
    \left(
        \log \GHpdf\bra{ \vyst ; \GHmRshrinkt, \GHvgamma, \GHnu}
        -
        \sum_{\dimi=1}^\dimN \log \GHpdf_{\dimi}\bra{\dysit ; \GHgammai, \GHnu}
    \right)
    ,
    \\
    \nonumber
    \GHmRshrinkt &= \diag(\mSigmashrinkt)^{-1/2}\,
    \mSigmashrinkt\, \diag(\mSigmashrinkt)^{-1/2},
    \qquad
    \mSigmashrinkt = \hat\mW\, \mLambdashrink\subt\, \hat\mW\trans,
    \\
    \nonumber
    \log \dmu\subitp 
    &= \left\{ \begin{array}{ll}
         (1-\dbi) \log \dmu\subi + \dbi\, \log \dmu\subit + \dai \dnablait,
         & \text{for } \dimi=1,\ldots,\dimNz, 
         \\
        \log  \dmu\subi, & \text{for } \dimi=\dimNz+1,\ldots,\dimN,
    \end{array} \right.
\end{align}
where $\vstaticpar$ gathers all the static parameters $\mA$, $\mB$, $\GHvgamma$, and $\GHnu$ of the model, and where we use the BIC to select the number of time-varying eigenvalues $\dimNz$, as explained in Section~\ref{subsec:nonlinear shrinkage}.
The static parameter vector $\vstaticpar$ is then estimated using Maximum Likelihood.

\section{Simulation study}
\label{sec:Sim}
To assess the performance of the high-dimensional dynamic copula model with spectral regularization in a controlled environment, we perform two simulation experiments.
In the first experiment, we focus on the performance of the spectral copula structure vis-\`a-vis a grouped factor copula structure \citep[as used in, e.g.,][]{Oh2017,Oh2018,Oh2023,Opschoor2020}. We investigate how the different copula structures and the shrinkage method behave under controlled deviations from a grouped factor structure.
In our second experiment, we focus on the parameter estimation performance of our new model.

\subsection{Experiment 1: spectral versus grouped copula structures} 
\label{subsec:SimStatic}

In both experiments, we specify the unconditional correlation matrix structure $\GHmR$ in our data generating process (dgp) as
\begin{equation}
    \label{eq:sim dependence structure}
    R_{i,j} =
    \frac{
        \SIMbetaM^2 +
        \SIMbetaGi^2 \krondelta_{\SIMgroupi,\SIMgroupj} +
        \SIMbetaI^2 \krondelta\subij+
        \SIMbetaC^2 e^{-|\SIMcountryi - \SIMcountryj|/2}
    }{
        \sqrt{
            \SIMbetaM^2 +
            \SIMbetaGi^2 +
            \SIMbetaC^2 +
            \SIMbetaI^2
        }\ \ 
        \sqrt{
            \SIMbetaM^2 +
            \SIMbetaGj^2 +
            \SIMbetaC^2 +
            \SIMbetaI^2
        }\ 
    }
    .
\end{equation}
Here, $\krondelta\subij$ is the Kronecker delta with $\krondelta\subij = 1$ if $\dimi=\dimj$, and zero else.
We can interpret this correlation structure as corresponding to a factor model of the form:
$ \SIMbetaM\,\factorMt + \SIMbetaGi\,\factorGit + \SIMbetaCi\,\factorCit + \SIMbetaI\,\factorepsit$,
where $\SIMbetaM$ is the loading coefficient of a common market factor $\factorMt$ to which all assets are exposed; 
$\SIMbetaGi$ is the factor loading for a group factor $\factorGit$ corresponding to group $\SIMgroupi$ of asset $\dimi$, e.g., an industry factor; 
$\SIMbetaC$ is the factor loading for the factor $\factorCit$, which defines, for instance, a country factor, i.e., a second grouping structure over and above the first (industry) grouping effect;
and finally $\SIMbetaI$ represents the size of the idiosyncratic noise component.
Higher values of $\SIMbetaI$ decrease the signal-to-noise ratio. 
In our example, the (industry) indicators $\SIMgroupi$ and (country) indicators $\SIMcountryi$ both range from 1 to 10, and the factors have zero means and are uncorrelated with each other.
The only exception is the country factor $\factorCit$, which is correlated between countries $\SIMcountryi$ with a correlation that varies with the distance $|C_i-C_j|$. 

If $\SIMbetaC=0$, the dgp collapses to a pure group structure as in \citet{Oh2017,Oh2018,Oh2023} and \citet{Opschoor2020}.
For $\SIMbetaC > 0$, however, the group structure is only approximate, which allows us to study the effect of such deviations on both our new copula structure as well as alternatives from the literature.
We consider $\dimN=100$ assets, where each asset corresponds to a unique pair $(\SIMgroupi,\SIMcountryi)$.
If $\SIMbetaC > 0$, the correlation structure can no longer be easily restored by a simple extension of the number of factors, while retaining the block-diagonality of the factor loading matrix and the orthogonality of the group factors; see Supplementary \ref{app:oh methodology}.

\begin{table} [t]
\caption{\label{tab:SimExpDynamic} Correlation properties of the empirical data set and the stylized dependence model. We show the properties of Gaussian rank correlations, such as the cross-sectional mean, standard deviation and the range over all correlations. We show also the mean, standard deviation and the range over two specific sectors, namely the industry sector with highest average correlation and lowest average correlation. }
\begin{center}
\begin{tabular}{ l ccc c ccc} 
\hline
 & \multicolumn{3}{c}{Empirical} & & \multicolumn{3}{c}{Simulation}\\
\cmidrule{2-4} \cmidrule{6-8} 
 & mean & std. dev. & range & & mean & std. dev. &  range  \\
\hline
All correlations    & 0.29 & 0.12 & $[0.00-0.78]$ &  & 0.29 & 0.15 & $[0.05-0.73]$ \\
Lowest correlated sector  & 0.30 & 0.12 & $[0.11-0.56]$ &  & 0.31 & 0.13 & $[0.14-0.58]$ \\
Highest correlated sector & 0.54 & 0.08 & $[0.41-0.71]$ &  & 0.59 & 0.07 & $[0.49-0.73]$ \\
\hline
\end{tabular}
\end{center}
\end{table}

We select the values of the parameters in the dgp to resemble the empirical complexity of the international stock market data as used in Section~\ref{sec:Emp}. 
We set $\SIMbetaI=1$, $\SIMbetaM = 0.75$ and let $\SIMbetaGi=1.75-0.15\dimi$ to have groups with different intragroup (industry) copula correlations. 
For $\SIMbetaC$, we consider a parameter range $\SIMbetaC \in \{0,0.75,1.5\}$. 
This means that country-effects are either fully absent, or are smaller than industry-effects, or have a comparable size. 
For $\SIMbetaC =0$, the grouped factor structure defined by the groups $\SIMgroupi$ is exact and all within-group correlations have the same value.
For $\SIMbetaC>0$, the within-group correlations start to differ and the group structure based only on $\SIMgroupi$ no longer captures the full heterogeneity in the dependence structure. 
Having considerable heterogeneity is in line with our empirical data study in Section~\ref{sec:Emp}.

Table \ref{tab:SimExpDynamic} shows various properties of unconditional Gaussian rank correlations for the empirical data from Section~\ref{sec:Emp}, and for simulated Gaussian copula data based on the stylized correlation matrix with $\SIMbetaC = 1.5$.
The two sets of correlations and their heterogeneity show similar properties between the simulated dgp and the empirical data.
The stylized parameters result in a cross-sectional average correlation of 0.29, which includes both intrasector correlations and cross-sector correlations. 
The correlations averaged over each sector in the simulation vary from 0.31 to 0.59. 
The intrasector correlations themselves vary from 0.14 to 0.58 within the lowest correlated sector and from 0.49 to 0.73 within the highest correlated sector.
Similar values and ranges are found for the data set used in Section~\ref{sec:Emp}.
The simulation setting thus mirrors the stylized facts in the empirical data quite well.

As our benchmark model in the simulations and in the empirical application later on, we use the dynamic factor copula approach of \citet{Opschoor2020} and \citet{Oh2023}.
We select the optimal number of clusters endogenously using the clustering methodology of \cite{Oh2023} based on their computer code.
See Supplementary \ref{app:additional simulations} for more details on the dependence structure imposed by their methodology.
We consider a $\dimN = 100$ dimensional time series and set the number of simulated observations to $\dimT \in \{250, 1000\}$.
We compute the models' performance metrics both in-sample and out-of-sample to illustrate the effect of over-fitting and of shrinkage.
On top of the $\dimT$ in-sample observations, we therefore generate an additional 1000 observations from the same dgp and re-calculate the log-likelihood without re-estimating the model's parameters to obtain the out-of-sample log-likelihood.
To simulate the data, we use a skewed $t$ with $\GHnu=25$ and $\GHvgamma = \gamma\,\viota$, where $\gamma = -0.25$ and $\viota$ is a vector of ones, as in \citet{Oh2023}.
These parameter values are similar to those obtained from empirical data.

\begin{table}[t]
\begin{center}
\caption{
    \label{tab:SimExpStatic} 
    Comparison of in-sample and out-of-sample log-likelihoods ($\ell_{in}$ and $\ell_{out}$) for different correlation structures, estimation methods and sample sizes $\dimT$ in $\dimN=100$ dimensions. 
    The true copula dependence structure $\GHmR$ is given in \eqref{eq:sim dependence structure} and either has a single clear grouped correlation structure (such as industry-only, $\SIMbetaC=0$), or a stylized, but realistic additional intra and inter-group correlation structure (e.g. country, $\SIMbetaC=0.75$ or 1.5).
    The $\dimT$ in-sample observations are used to estimate the models.
    An additional $\dimT$ simulated out-of-sample observations are used to compute $\ell_{out}$. 
    In the dgp, data are generated using the skew $t$ copula with $\GHnu=25$ and $\GHvgamma = -0.25\,\viota$.
    Best performing models per combination of dependence structure and sample size (row-wise) are bolded.
}
\begin{tabular}{cc cc c cc c cc c cc} 
\hline
  $\SIMbetaC$  & $\dimT$ & \multicolumn{2}{c}{True } &  & \multicolumn{2}{c}{Regularized} & & \multicolumn{2}{c}{Sample} &  &   \multicolumn{2}{c}{Factor} \\
  \cmidrule{3-4} \cmidrule{6-7} \cmidrule{9-10} \cmidrule{12-13}
 & & $\ell_{in}$ & $\ell_{out}$ & & $\ell_{in}$  &  $\ell_{out}$ & & $\ell_{in}$  &  $\ell_{out}$ & & $\ell_{in}$  &  $\ell_{out}$\\
\midrule
0    & $1000$ & 35,168 & 36,206 & & 36,899 & 34,798 & & 37,563 &  33,366 &  & 34,811 & $\bm{35,791}$   \\
0    & $250$  & 8,562  & 8,577 & & 9,925  &  7,694 & & 11,496 & 4,380 & &  8,555  &  $\bm{8,507}$ \\  
0.75 & $1000$ & 39,392 & 39,665 & & 41,062 & $\bm{38,135}$ & & 41,800 & 36,805 &  & 30,922 & 31,304 \\
0.75 & 250    & 9,422  &  9,414 & & 10,903 & $\bm{8,283}$ & & 12,399 & 5,168 & &  7,402  &  7,313  \\
1.5  & $1000$ & 55,693 & 55,748 & & 57,185 & $\bm{54,459}$ & & 58,132 & 52,906 &  & 33,943 &  33,439 \\
1.5  & 250    & 13,611 & 13,529 & & 15,051 & $\bm{12,464}$ & & 16,601 & 9,295 & & 8,343  &  8,120  \\
\hline
\end{tabular}
\end{center}
\end{table}

Table~\ref{tab:SimExpStatic} presents the results.
There are three main takeaways.
First, if there is a true factor structure ($\SIMbetaC=0$), then the factor cluster copula model of \citet{Oh2023}, labeled Factor, performs best based on the out-of-sample log-likelihood $\ell_{out}$, with the regularized spectral model in second place.
This is encouraging, as the out-of-sample performance loss for the spectral method vis-\`a-vis a correctly specified more parsimonious model is limited.
To appreciate this, we note that the spectral dynamic copula model is agnostic about the true model structure and therefore still requires the estimation of 4950 different correlation parameters.

Second, we see that MLE combined with the sample correlation matrix (in the column Sample) suffers from the curse of dimensionality for large correlation matrices, as expected \citep{Ledoit2022a}.
The out-of-sample performance of the MLE is significantly worse than its in-sample performance. 
The biases in the spectrum become more pronounced if the number of in-sample data points decreases to, e.g., $\dimT=250$. 
The problem of overfitting is resolved either when a parsimonious model structure is imposed through a cluster factor structure (in the column Factor) or when we use the quadratic shrinkage approach of \citet{Ledoit2022b} (in the column Regularized).

Third, when we depart from the true group structure, the results for $\SIMbetaC=0.75$ and 1.5 show that the spectral copula approach with regularization performs best by a wide margin.
Again, when shrinkage is not applied, we see similar biases as before.
Also, the effect of shrinkage becomes larger as the ratio $\dimN/\dimT$ increases.
We also see that the differences in out-of-sample log-likelihood performance between the clustering and the regularized approach is large for $\SIMbetaC = 0.75$ and 1.5. 
This holds even if one allows for an endogenous choice of the number of groups and asset group assignments as in \citet{Oh2023}.
In the stylized correlation matrix from Eq.~\eqref{eq:R parameterization}, there are variations in correlations both within groups and across groups, while the cluster factor approach imposes that all correlations for stocks belonging to a group are the same.
Moreover, the cluster factor structure does not span the full space of correlation matrices (see Supplementary \ref{app:additional simulations}).
As a result, the cluster factor structure can miss out on important forms of heterogeneity in the data, both in the simulation and the empirical study.

\subsection{Experiment 2: quality of point and path estimates} 
\label{subsec:SimDynamic}

\begin{table} [t]\centering
\caption{\label{tabSimExpDynamic} Simulation results for the copula parameter estimates based on the dynamic skew $t$ copula as dgp with the dependence structure from \eqref{eq:sim dependence structure}. The sample size is $\dimT=1000$ in $\dimN=100$ dimensions.
Sample ML uses the sample spectrum of $\hat\GHmR$ in Eq.~\eqref{eq:target omega2}, while Regularized ML uses the quadratic shrinkage formula of \citet{Ledoit2022b}. The simulations are repeated 100 times. The mean and the standard deviation of the estimates are reported. 
}
\begin{tabular}{ l r c r r c r r } 
\hline
 &  &  & \multicolumn{2}{c}{Regularized ML} &  & \multicolumn{2}{c}{Sample ML} \\
\cmidrule{4-5} \cmidrule{7-8} 
Parameter   & True & & Mean & Std. dev. & & Mean & Std. dev.  \\
\midrule
$\lambda_1$     & 29.8  &  & 29.4 & (1.8)    &  & 29.7  & (1.7)        \\
$\lambda_2$     & 10.7  &  & 10.7 & (0.8)    &  & 10.8  & (0.7)        \\
$\lambda_{99}$ & 0.160 &  & 0.171 & (0.008) &  & 0.101 & (0.003)     \\
$\lambda_{100}$ & 0.160 &  & 0.164 & (0.009) &  & 0.097 & (0.003)     \\
$\da_1$     & 0.10  &  & 0.09 & (0.01)   &  & 0.09  & (0.01)    \\
$\db_1$     & 0.90  &  & 0.89  & (0.03)  &  & 0.89  & (0.03)    \\ 
$\da_2$     & 0.10  &  & 0.09 & (0.02)   &  & 0.08  & (0.02)   \\
$\db_2$     & 0.90  &  & 0.88  & (0.06)  &  & 0.88  & (0.07)    \\ 
$\GHnu$     & 25.0  &  & 24.6  & (1.8)   &  & 29.2  & (2.2)     \\
$\gamma$    & -0.25 &  & -0.28  & (0.07) &  & -0.31 & (0.08)     \\
\hline
\end{tabular}
\end{table}

So far, the simulation study concentrated on the unconditional dependence structure of the copula and the performance of the shrinkage procedure. 
Next, we consider the quality of the point estimates of the copula parameters, the selection of the number of dynamic components, and the path estimates of the dynamic eigenvalues.
We take the score-driven dynamic spectral copula model from Section~\ref{sec:Def} as our dgp.
The stylized correlation matrix from \eqref{eq:sim dependence structure} is again used with $\SIMbetaC=1.5$.
The other parameter values (see Table~\ref{tab:SimExpDynamic}) are chosen to be similar to the empirical results from Section \ref{sec:Emp}.
The first 2 eigenvalues in the dgp are dynamic, while the remaining 98 eigenvalues in the dgp are static, also in line with the empirical results.
We estimate the static parameters using the Sample ML as well as the Regularized ML procedure as described in Section~\ref{subsec:nonlinear shrinkage}.
The estimation results can be found in Table~\ref{tabSimExpDynamic}.

The results in Table~\ref{tabSimExpDynamic} confirm that the Sample ML approach suffers from biases in the estimation of the eigenvalues at the lower end of the spectrum, i.e., for small eigenvalues.
At the upper end of the spectrum, i.e., for the large eigenvalues, the effect of shrinkage is less pronounced.
This means that the Regularized ML estimator successfully removes the biases at the low end of the spectrum, while leaving `spiked' eigenvalues intact.
We note that also all other parameters, such as $\dai$, $\dbi$, $\GHnu$, and $\GHvgamma=\gamma\,\viota$, are estimated accurately.

\begin{figure}[t]
\centering
\includegraphics[width=1.0\columnwidth]{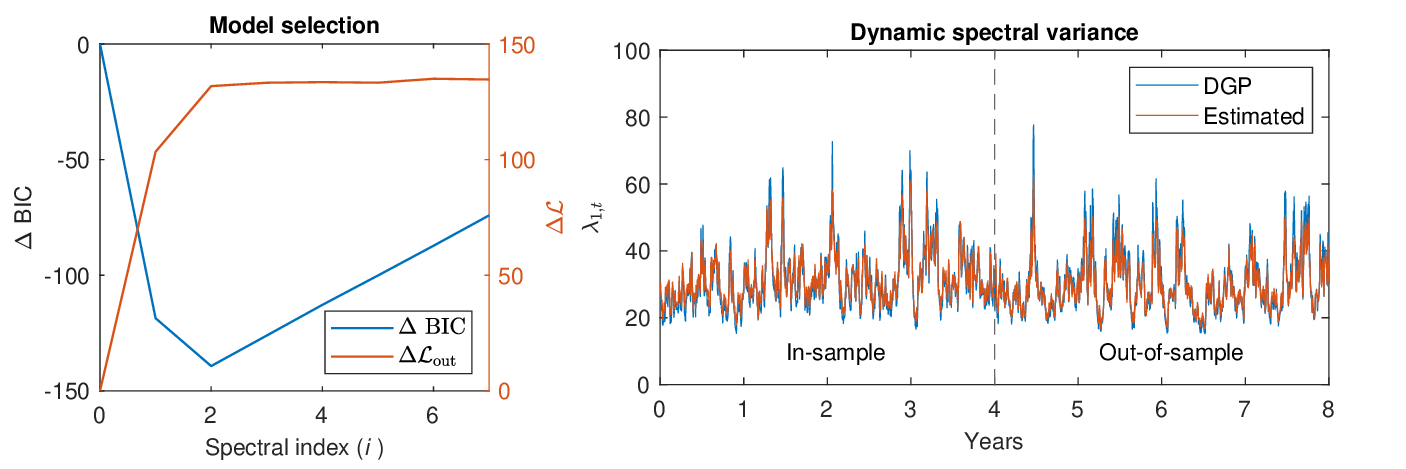}
\caption{Performance of the skewed $t$ copula with regularized spectral dynamics. The left plot shows changes in the (in-sample) BIC of a model with static eigenvalues versus a model with eigenvalues $1,\ldots,\dimi$ being dynamic, as a function of the spectral index $i$.
The BIC is minimal for $i=2$, which is also the true number of dynamic eigenvalues in the dgp. The average result of 10 Monte Carlo simulations is shown. The right-hand plot shows the dynamics of the first eigenvalue for a single MC simulation, where the true dynamics of the dgp is compared with the predicted dynamics from the estimated model. The period after 4 years is an out-of-sample forecast. }\label{sim:figEigValDyn}
\end{figure}

Figure~\ref{sim:figEigValDyn} presents the results for the model selection procedure to determine the number of dynamic eigenvalues.
In each Monte Carlo simulation and for $\dimi=1,\ldots,7$, we successively estimate models with eigenvalues $1,\ldots,\dimi$ as dynamic, and eigenvalues $\dimi+1,\ldots,\dimN$ as static.
We compute the in-sample BIC decrease with respect to a fully static model ($\dimi=0$), as well as its out-of-sample log-likelihood increase. 

The left-hand panel in Figure~\ref{sim:figEigValDyn} shows that the average $\Delta$BIC curve has its minimum at the correct number of $\dimN_0=2$ dynamic components.
This analysis validates our choice to use BIC for selecting the number of dynamic eigenvalues in the empirical study of Section~\ref{sec:Emp}. 
From the out-of-sample log-likelihood increases (right-hand axis in Figure~\ref{sim:figEigValDyn}), we see that it is particularly important to include the dynamics of the first eigenvalue, since it has the largest impact.
For spectral indices above $\dimi=2$, there are no further increases in log-likelihood if we make these eigenvalues dynamic, and the BIC grows linearly in the number of parameters.
The stable out-of-sample log-likelihood suggests that the risk in overfitting the number of dynamic components is relatively low, since overstating the number of dynamic eigenvalues does not have a negative effect on the out-of-sample log-likelihood.

The right-hand panel in Figure~\ref{sim:figEigValDyn} shows a simulated path of the first dynamic eigenvalue with the dgp and the estimated path from the dynamic copula model.
This shows that the eigenvalue dynamics are captured accurately.
An additional simulation experiment in Supplementary~\ref{app:sim misspec} confirms that the dynamic score-driven model can still recover the true, unobserved eigenvalue dynamics even if it is mis-specified, in line with theoretical consistency results in \citet{beutnerlinlucas2023}.

\section{Empirical study}
\label{sec:Emp}

\subsection{Data}
\label{subsec:data}

In our empirical study, we consider daily log return data for 100 stocks from 10 different industry sectors and 10 different European countries over a period of 10 years. 
The data are obtained from LSEG (formerly known as Refinitiv), so that our industry sectors are based on The Refinitiv Business Classification (TRBC), which distinguishes between Financials (FI), Industrials (IN), Technology (TE), Basic Materials (BM), Consumer Cyclicals (CC), Consumer Non-Cyclicals (CN), Utilities (UT), Health Care (HC), Energy (EN) and Real Estate (RE). 
We select stocks that are issued and traded in 10 different European countries: Germany (DE), United Kingdom (UK), France (FR), Spain (ES), Italy (IT), Sweden (SE), Norway (NO), The Netherlands (NL), Belgium (BE) and Switzerland (CH). For each sector and country combination, we select stocks with the largest market capitalizations after the end of the observation period, while also requiring that each stock is fully observed over the 10 year period from January 2015 to 31 December 2024.
We have a few (8) missing country-industry combinations that do not satisfy our criteria, e.g., because they are not observed over the full sample period.
In such cases, we take the stock from another country in that same industry (with second highest market capitalization).
We are able to do so in such a way that we end up with each of the 10 countries and the 10 sectors occurring precisely 10 times, where some country-sector combinations will be missing, while other combinations occur twice.
Table \ref{tabIndices100} of Supplementary~\ref{app:additional empirics} lists the full set of all 100 stocks across countries and sectors. 

Thus far, high-dimensional copula studies have mainly focused on stocks from a single country, such as the US \citep[see, e.g.,][]{Oh2023}. 
The large number of possible country and sector combinations increases the complexity of the dependence structure in the current data set and poses challenges to standard factor clustering algorithms.
As a result, copulas that do not impose strong restrictions on the correlation structure can achieve substantial performance gains.

\begin{table} [t]\centering
\caption{\label{tabEmpiricalMarginal} Summary statistics of univariate times series. Panel A presents the unconditional mean, standard deviation, skewness and kurtosis of the daily log-returns. Panel B shows the estimated parameters of the marginal AR(1)-GARCH(1,1) models for the univariate time series,
$
r_{i,t}= \delta_{i} + \phi_{i} r_{i,t-1}+ \epsilon_{i,t}$ and $\sigma^2_{i,t} = \omega_{i} + \alpha_{i} \epsilon^2_{i,t-1}+ \beta_{i} \sigma^2_{i,t-1}$, estimated by QMLE.
Panel C shows the correlations of the devolatilized returns $\dyit = \epsilon_{i,t}/\sigma_{i,t}$. In all cases the mean and the range over the cross-sectional dimension of 100 stocks is given.}
\begin{tabular}{ l p{0.1cm} cc c l cc} 
\hline
Cross section & & mean & range & & & mean & range  \\
 \hline
\multicolumn{8}{l}{Panel A: log-returns $r_{i,t}$}\\
\hline 
Mean & & 0.000 & $[-0.001,\,0.001]$ & & Skewness  & $-0.406$ & $[-1.82,\,1.25]$ \\
Std. dev.  & & 0.018  & $[0.011,\,0.030]$ & & Kurtosis & 12.95 & $[5.72,\,35.25]$ \\
\hline
\multicolumn{8}{l}{Panel B: univariate AR(1)-GARCH(1,1) parameter estimation results}\\
\hline 
$\sqrt{\omega_{i}}$ & & 0.004 & $[0.001,\,0.011]$  & & $\delta_{i}$  & 0.000 & $[-0.001,\,0.001]$ \\
$\alpha_{i}$ & & 0.090  & $[0.013,\,0.317]$ & & $\phi_{i}$ &-0.023 & $[-0.195,\, 0.061]$ \\
$\beta_{i}$  & & 0.846 & $[0.548,\,0.979]$  & &  &  &  \\
\hline
\multicolumn{8}{l}{Panel C: Average cross-sectional correlations devolatilized residuals $\dyit = \epsilon_{i,t}/\sigma_{i,t}$}\\
\hline
$\rho_{i,j}$ & & 0.275  & $[0.004,\,0.786]$ & & & & \\
\hline
\end{tabular}
\end{table}

We merge the time series of the log-returns from different countries based on common trading days.
This results in a data set with $\dimT = 2,425$ daily returns $r_{i,t}$ for each asset $i=1,\ldots,\dimN$. 
Summary statistics of the data are presented in Panel A of Table \ref{tabEmpiricalMarginal}, confirming standard stylized facts such as unconditional left-skewness and substantial excess kurtosis of the raw log-returns. 

\begin{figure}[!t]
\centering
\includegraphics[width=1.0\columnwidth]{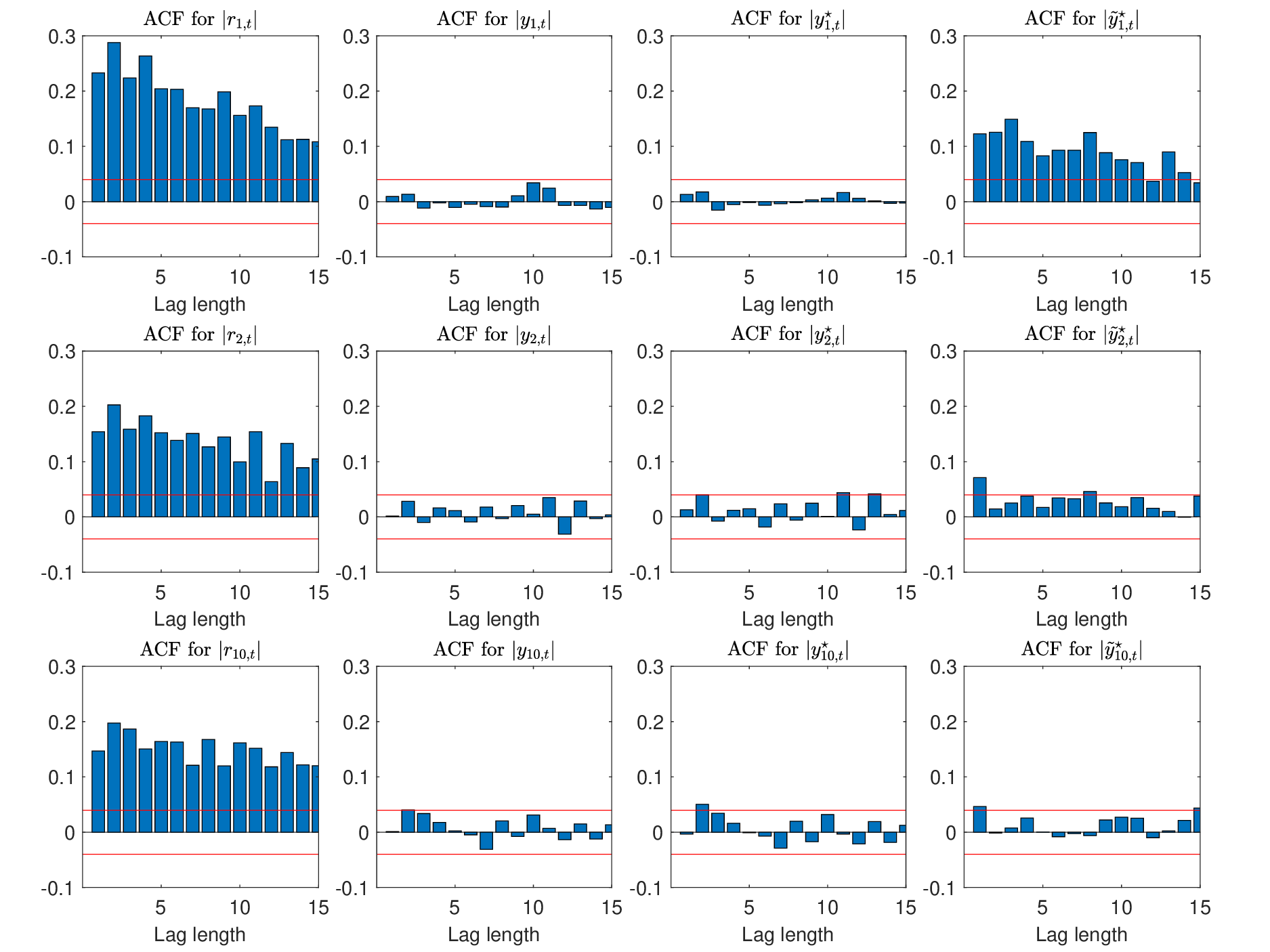}
\caption{Autocorrelation functions for log-returns $|r_{i,\dimt}|$, devolatilized residuals $|\dyit|$,  $|\dysit|= |\GHcdf^{-1}(\dPITit)|$ and spectral projections $|\dystildeit|$ for $i=1,2,10$. The Gaussian copula ($\GHgamma=\vzeroN$ and $\GHnu^{-1}=0$) is used to determine $\dysit$ and $\dystildeit$.}\label{figAcf}
\end{figure}

We filter each time series $r_{i,t}$ using a standard AR(1)-GARCH(1,1) filter
and proceed the analysis with the devolatilized residuals $\dyit=\epsilon_{i,t}/\sigma_{i,t}$. 
Panel B of Table \ref{tabEmpiricalMarginal} summarizes the univariate estimation results.
The first two columns of panels in Fig. \ref{figAcf} show the autocorrelation functions for absolute log-returns $|r_{i,t}|$ and for absolute devolatilized returns $|\dyit|$ for 3 (arbitrarily chosen) stocks. 
As expected, we observe strong serial correlation for the $|r_{i,t}|$, which indicates clear volatility clustering effects.
After applying the univariate AR(1)-GARCH(1,1) filters, the standardized absolute residuals $|\dyit|$ no longer indicate any substantial volatility clustering.
We take these standardized residuals $\dyit$ as input for the Probability Integral Transformations (PITs) using the rank-based transforms $u_{i,t} = \text{rank}(\dyit)/(T+1/2)$ as motivated in Section~\ref{subsec:parameter estimation}.
Our final sample consists of $T = 2,425$ pseudo-copula observations in $\dimN=100$ dimensions. 
We use the first half of the data for estimation, and the second half for our out-of-sample performance evaluation.

Interestingly, we can use the pseudo-copula observations $\dysit = \GHcdf\subi(u\subit)$ to explore whether there will be time-variation in the eigenvalues of the copula dependence matrix.
The third column in Figure~\ref{figAcf} provides the autocorrelation function of the absolute pseudo-copula observations $|\dysit|$.
Again, we see no substantial signal of volatility clustering in the values of $\dysit$.
However, if we compute the eigenvalue-eigenvector decomposition of the sample correlation matrix of $\vyt$ and use it to compute initial estimates of the spectral projections $\dystildeit$, the last column of Figure~\ref{figAcf} clearly shows volatility clustering effects for the first spectral projection, which is indicative of time-variation in $\dlambda_{1,\dimt}$.
As another example, also the second spectral projection is shown, for which the time-varying volatility, i.e., a time-varying $\dlambda_{2,\dimt}$, is less strong.
The further down in the spectrum we consider the spectral projections, the less evidence we find for a time-varying $\dlambdait$.
For instance, for $\dimi=10$ the lower-right autocorrelation function largely remains within the confidence band.
This indicates that a modeling approach with a limited number of time-varying $\dlambdait$s for the copula dependence parameters is both parsimonious and congruent with the financial data at hand.

Before presenting the estimation results for the dynamic spectral regularized copula specification, we note that our analysis differs conceptually from existing (spectral) GARCH models in the literature, such as the orthogonal GARCH or the $\lambda$ GARCH models \citep[e.g.,][]{hetland2023dynamic}.
The latter do not explicitly distinguish between the marginal and the copula time series.
As a result, the conditional volatility effects in spectral directions in those models could stem from marginal volatility increases as well as from increased correlation effects. 
In our framework, by contrast, a larger conditional variance $\lambda_{1,t}$ solely reflects the dynamics of the copula, allowing us to answer whether strongly dependent market movements (as measured by the first spectral projections) are followed by periods of higher dependence. 

\subsection{Estimation results}

Given that the time-variation in $\dlambdait$ is concentrated in the first few spectral projections according to the autocorrelograms in Figure~\ref{figAcf}, we use a BIC guided selection procedure to determine the final number of dynamic components.
We start from the static skew $t$ copula, after which we make the first eigenvalue dynamic.
We report the change in (unregularized) BIC relative to the static model. 
After that, we also make the second value dynamic and report the change in BIC relative to the static model, and so on.
We do not apply shrinkage at this stage.
Figure~\ref{figEigValDyn} shows the changes in the BIC vis-\`a-vis the static model when adding the dynamic eigenvalues for the skewed $t$ copula one by one for $\dimi=1,...,7$.
Clearly, the largest gain is obtained by allowing the first $\dlambdait$ to be dynamic, followed by a smaller improvement for $\dimi=2$.
The BIC is at its minimum for $\dimi=2$ dynamic components, which therefore is the value we use in the remainder of the analysis.
We use the same number of 2 dynamic eigenvalues for the three different copula densities, endowing them with the  score-driven dynamics from Section~\ref{subsec:dynamics and shrinkage}.

\begin{table}[t]
\caption{\label{tabLoglCase} 
Performance of various static and dynamic copula specifications with either a factor structure or a regularized spectral structure for the correlation matrix.  We consider 100 European stocks from 10 different countries and industry sectors ($d=100$) with 10 years of daily data. The first 5 years are used for estimation and the last 5 years for out-of-sample performance. 
Panel A shows the in-and out-of-sample log-likelihood for dynamic copula specifications. 
Panel B shows the same results based on static copula specifications.
The best performing copula specification is indicated in bold, which is the dynamic skew $t$ copula with non-linear shrinkage.
}
\begin{tabular}{l c c c c c c c c c c c} 
\hline
 & \multicolumn{3}{c}{Regularized} & & \multicolumn{3}{c}{Sample} & & \multicolumn{3}{c}{Factor}  \\
 \cmidrule{2-4} \cmidrule{6-8} \cmidrule{10-12}
 & \it{G} & $t_d$ &  skew $t_d$ &  & \it{G} &  $t_d$ &  skew $t_d$ &  & \it{G} &  $t_d$ &  skew $t_d$  \\ 
\hline
\multicolumn{10}{l}{Panel A: dynamic copula likelihoods}\\
\hline
$\ell_{in}$ & 38,068 & 38,537 & 38,574 & & 38,220 &  39,012 & 39,045 & &  28,376 &  29,641 & 29,686 \\
$\ell_{out}$ & 27,882 & 29,725 & $\bm{29,752}$ & & 27,139 & 28,705 & 28,733 & & 24,882 &  26,531 & 26,555\\
\hline
\multicolumn{10}{l}{Panel B: static copula likelihoods}\\
\hline
$\ell_{in}$ & 37,397  & 38,058 & 38,104 & & 37,664 & 38,650 & 38,691 & &  28,088 &  29,481 & 29,529 \\
$\ell_{out}$ & 26,769 & 29,529 & 29,549 & & 25,513 & 28,230 & 28,253 & & 23,605 &  25,696 & 25,727 \\
\hline
\end{tabular}
\end{table}

Table \ref{tabLoglCase} shows the performance for the three different approaches (Regularized spectral copula, Sample-based spectral copula, and the Factor copula with clustering) and three different choices for the copula density (Gaussian, Student's $t$, and the skewed $t$).
We estimate the factor model with optimal cluster assignment in 100 dimensions using the algorithm of \cite{Oh2023}, where 17 clusters turn out to be optimal in terms of BIC.
Panels A and B of Table \ref{tabLoglCase} show the in-sample and out-of-sample log-likelihood values for the different dynamic and static models, respectively.
First, we find that the spectral dynamic copula models (panel A) perform significantly better than the static copula models (panel B), both in-sample and out-of-sample.
Modeling the dynamics of the dependence structure thus significantly improves the fit and predictive power of the models, even in the current parsimonious setting where only 2 out of the 100 eigenvalues are dynamic.
Second, the dynamic factor copula specifications with endogenous clustering perform significantly worse compared to the regularized dynamic spectral copula specifications.
This is mainly due to their more restrictive dependence structure compared to the spectral copula specification.
Third, as expected and as mentioned earlier, the $t$ copula performs substantially better than its Gaussian counterpart.
Performance differences between the symmetric and the skew $t$ copulas, by contrast, are modest, though there is a slight increase in the log-likelihood, both in-sample and out-of-sample.
Finally, non-linear shrinkage of the unconditional eigenvalues leads to significant improvements of $+1000$ points in the out-of-sample log-likelihood compared to the sample eigenvalues.
Spectral dynamics and regularization thus emerge as the most important aspects in modeling the copula dependence spectrum for large $\dimN$ and $\dimT$.

In Table \ref{tabEstCase}, we show the parameter estimates for the two dynamic spectral indices of the best performing dynamic skew $t$ copula with regularization.
To obtain confidence intervals for the estimates, the block bootstrap method was used on pseudo-copula observations $u_{i,t}$ using 20 trading days as block length and 200 bootstrap samples.
For both eigenvalues, the dynamics are highly persistent with $\db_1$ and $\db_2$ equal to 0.9 or higher.  
The value of $a_1$ can be accurately estimated and appears clearly significant, signaling a time-varying dependence structure. 
The value of $a_2$ can be less accurately estimated and the corresponding improvement in log-likelihood is smaller, but it is still yields a lower BIC and also the out-of-sample log-likelihood improves.
Such observations are also in line with Figure~\ref{figAcf}.
The estimated value for the skewness parameter $\GHgamma$ is significantly negative, while the estimated tail parameter $\GHnu$ is fairly large with values around 45. Still, the improvement in log-likelihood is substantial when allowing for tail-dependence, both in sample and out-of-sample, as can be seen from Table \ref{tabEstCase} when comparing the (skewed) $t$ copula with its Gaussian counterpart.

\begin{table}[t]\centering
\caption{\label{tabEstCase} 
Estimation results for the skew $t$ copula with spectral dynamics based on 100 European stocks and 5 years of historic data. 
The 90\% confidence intervals are determined with the block bootstrap method on the pseudo-copula observations $u_{i,t}$ using 20 trading days as block length. 
}
\begin{tabular}{lcc ccc ccc} 
\hline
Par. & Est. & 90\% CI & Par. & Est. & 90\% CI & Par. & Est. & 90\% CI \\
\hline
$\dmu_1$\rule{0ex}{2.5ex}  & 30.6 & [26.2,\,\,35.5] & $a_{1}$ & 0.06 & [0.05,\,\,0.07] & $\db_1$ & 0.90 & [0.86,\,\,0.94] \\ 
$\dmu_2$  &  5.8 & [ 5.3,\,\, 6.6] & $\da_2$ & 0.06 & [0.03,\,\,0.21] & $\db_2$ & 0.97 & [0.26,\,\,0.99] \\
$\nu$    & 44.1 & [41.7,\,\,54.8]  & $\gamma$ & -0.37 & [-0.47,\,\,-0.21] \\
\hline
\end{tabular}
\end{table}

\subsection{Eigenvalues and eigenvectors}

\begin{figure}[t]
\centering
\includegraphics[width=1.0\columnwidth]{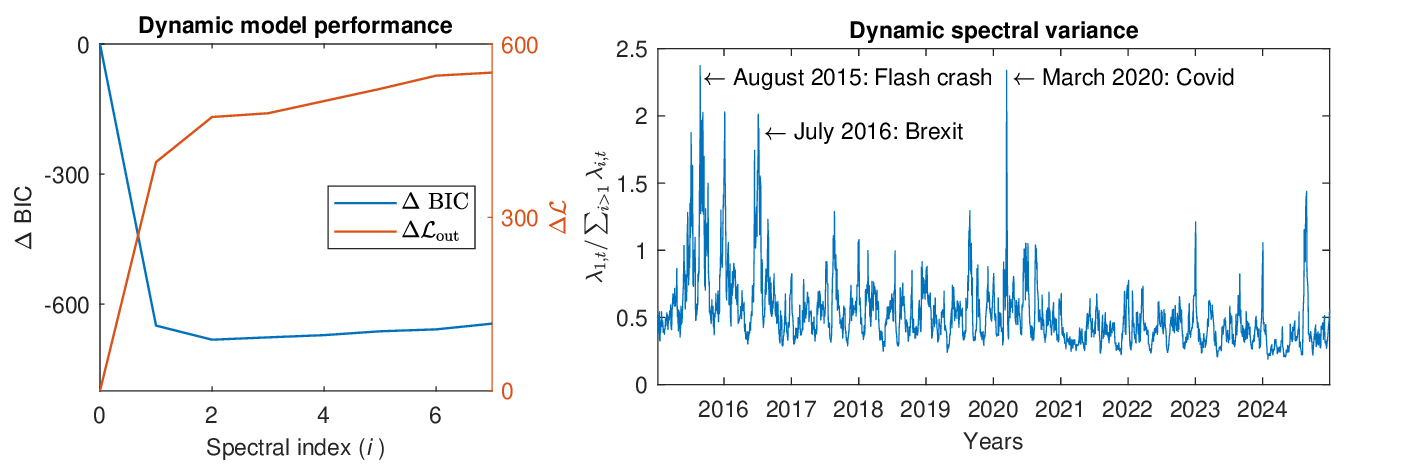}
\caption{Spectral dynamics of the first spectral eigenvalue, where the ratio of the first eigenvalue and the sum of the higher eigenvalues is plotted over time. Under normal market circumstances, the first eigenvalue explains about 30\% of the spectral variance, leading to a ratio below 1/2. When the ratio increases, it signals enhanced spectral variance in the parallel direction (first eigenvector) compared to the total spectral variance in the orthogonal directions, indicating less diversification and higher systemic risk or return. The three most pronounced events are given an economic interpretation. The period from January 2020 to December 2024 represents an out-of-sample forecast. }\label{figEigValDyn}
\end{figure}

\begin{figure}[tb]
\centering
\includegraphics[width=1.0\columnwidth]{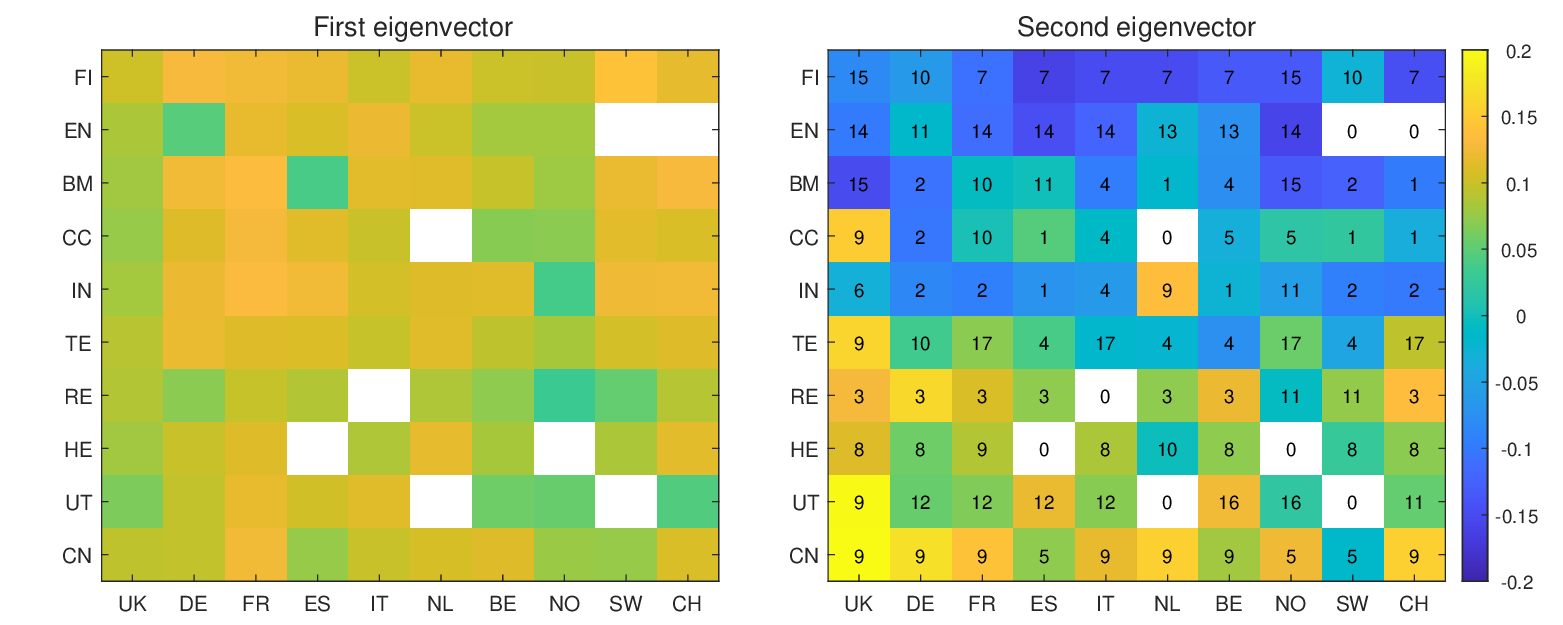}
\includegraphics[width=1.0\columnwidth]{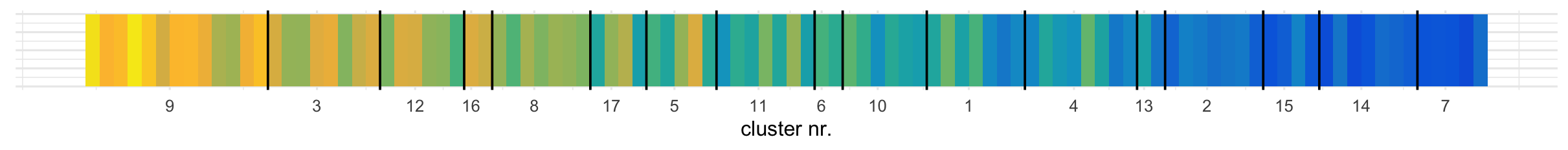}
\caption{Heatmap of estimated eigenvectors weights $\hat{w}_{i,j}$ of the unconditional copula correlation matrix $\hat\GHmR$ from Eq.~\eqref{eq:target omega2}. The first two eigenvectors are shown ($j=1,2$). The stocks $i$ are ordered in terms of countries along the $x$-axis and in terms of sectors along the $y$-axis. The white color indicates that the country-sector combination is absent in the data set. In case of two stocks per country-sector combination, the result for the first stock from Table \ref{tabIndices100} is shown. The first eigenvector has positive weights for all stocks, corresponding to a parallel market movement. We compare the second eigenvector with optimal cluster assigments using \citet{Oh2023} giving 17 clusters, indicated by the numbers in each sector-country cell. The colors relate to the value of the corresponding element of the eigenvectors in the spectral copula specification.
The bottom panel contains the same information as the right-hand panel, but the stocks are ordered per cluster (in order of their average value of the second eigenvector elements).
}\label{figEigVec}
\end{figure}

To conclude the empirical analysis, we zoom in on several aspects of the estimation results, such as the eigenvalue dynamics and its relation to financial crises, as well as the interpretation of the first eigenvectors.
We use the best performing copula model, namely the skew $t_d$ copula with regularized spectral dynamics. 
In the right plot of Fig.~\ref{figEigValDyn}, we show the dynamics of the first eigenvalue $\lambda_{1,t}$, where we plot the ratio of the first eigenvalue to the sum of the smaller eigenvalues ($i\leq 2$) over time. 
Under normal market circumstances, the first eigenvalue explains about 30\% of the spectral variance, leading to a ratio around 30/70, which is below 0.5. 
There is strong serial dependence in the first eigenvalue, in line with the earlier result in the right-hand panels in Figure~\ref{figAcf} and Panel A in Table~\ref{tabEstCase}. 
As is seen in Figure~\ref{figEigVec}, the first eigenvector takes a more or less equally weighted position in each of the assets, thus reflecting a parallel movement in the market.
Increases in the first eigenvalue compared to the remaining ones therefore result in correlation matrices with less diversification potential, leading to periods with higher systemic risk.
The right-hand panel in Figure~\ref{figEigValDyn} reveals that such parallel-to-orthogonal variance ratio increases can reach up to a factor two or higher during crisis periods, implying the first eigenvalue explains up to 70\% of the spectral variance during such times.
During these crisis events, the cross-sectional average correlation $\bar{\bm{R}}_t$ peaks at 0.65, while the unconditional average is only 0.30.
The three most pronounced events in the time period of Figure~\ref{figEigValDyn} are the flash crash in 2015, the Brexit in 2016 and the Covid pandemic in 2020, which corroborate the lack of diversification potential during such periods.

In Figure~\ref{figEigVec}, we analyze the eigenvectors corresponding to the largest two dynamic eigenvalues in more detail and compare them to the cluster assignments based on \citet{Oh2023}.
This is an extention to \citet{Gubbels2025}, who perform the eigenvector analysis solely for country-effects in a static spectral copula context.
The left panel shows the first eigenvector as a heatmap.
All elements are positive and of roughly equal magnitude, so that the first eigenvector represents the collective movement of the European stock market as a whole.
This is in line with the market factor interpretation of the first factor by \citet{Oh2023}: the estimated intercepts of the factor loadings for the optimal number (using the BIC) of 17 clusters are all positive, see Table \ref{tabEstClus100D} of Supplementary~\ref{app:additional empirics}.
Moreover, we also see from Figure~\ref{figEigVec} that the stocks with lowest weight in the first eigenvector (green color in the left panel) all correspond to cluster 11, which has the lowest market co-movement in Table~\ref{tabEstClus100D}.
More generally, the inner product of the first eigenvectors of the unconditional correlation matrix from both approaches is 0.994.

To further compare the spectral eigenvectors with the optimal cluster assignments based on the methodology of \cite{Oh2023}, we label each of the assets by their cluster number 1,\ldots,17, where 17 denotes the optimal number of clusters in terms of the BIC.
The right-hand panel in Figure~\ref{figEigVec} shows the heatmap of the second eigenvector, which is orthogonal to the first eigenvector.
It represents the most important cross-sectional direction for diversification in the European stock market.
The eigenvector mainly captures the diversification between different sectors, where the financial, energy and basic material sectors have opposite sign to the real estate, health, utility and consumer non-cyclical sectors.
From Table~\ref{tabEstClus100D} and Figure~\ref{fig:omega OhPatton vs eigvec 2}, we see that clusters 3, 7, 12, 14 have the largest absolute cluster loadings $\omega^{C}_{i}$) and represent the sectors real estate, financials, utilities and energy.
For these clusters, the corresponding eigenvector weights are most negative for sectors 7 and 14, and most positive for sectors 3 and 12.
This means that there is alignment between the two models in identifying co-moving stocks. 
Although weaker than sector effects, we also observe country effects in the second eigenvector.
The colors that correspond to Norway are more aligned with Sweden than with the UK, for example. 
This shows that in our heterogeneous data set both industry sectors and countries have intertwined effects.
To further illustrate the alignment between cluster assignments and the second eigenvector, the bottom panel in Figure~\ref{figEigVec} shows the heatmap of the weights grouped per cluster and sorted by their (cluster) average weight.
The cluster factor selects clusters with similar second eigenvalue weights in the spectral decomposition.
The main difference between the two approaches, however, lies in the handling of the between-cluster correlations: in the cluster factor approach with its block-diagonal cluster loading matrix (see Supplementary~\ref{app:oh methodology}) the off-diagonal factor loadings are restricted to zero, whereas no such restriction applies in the spectral approach.
This is the primary cause why the spectral approach obtains a better fit to the data.

\begin{figure}[t]
\centering
\includegraphics[width=1.0\columnwidth]{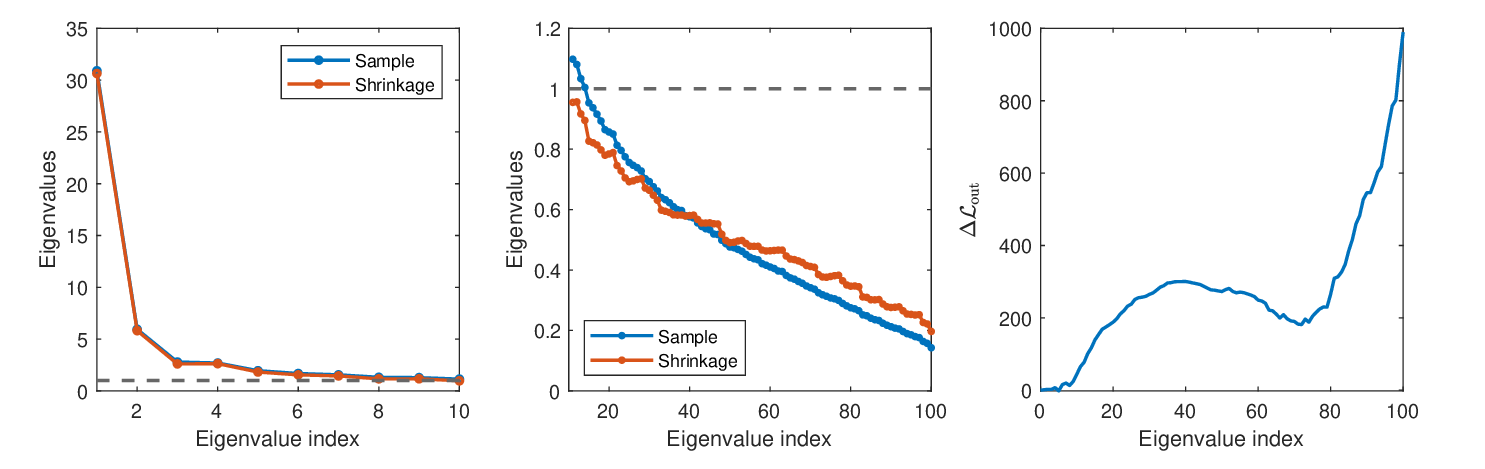}
\caption{Estimated eigenvalues excluding and including shrinkage, $\hat{\lambda}_{i}$ and $\hat{\mu}_{i}$, of the unconditional copula correlation matrix for the dynamic skew $t$ copula. The sample eigenvalues are ordered from high to low. The left plot shows the first 10 eigenvalues and the middle plot the other 90 eigenvalues. The sample eigenvalues are shown in blue, while the eigenvalues after applying shrinkage are shown in red. The right plot shows the out-of-sample log-likelihood improvement when the first $i$ sample eigenvalues are replaced with shrinkage eigenvalues. The largest improvements stem from the highest spectral indices, whose sample eigenvalues are most severely biased. }\label{figShrinkage}
\end{figure}

As final analysis, we zoom in on the effect of shrinkage on the empirical results.
When $\dimN = 100$, the concentration ratio $\dimN/\dimT$ is 8\%, since we use $\dimT=1,213$ in-sample observations for estimation.
Though modest, the biasing effect on the spectrum is clearly present, as we have seen in the earlier results.
To understand how the quadratic shrinkage procedure contributes to the fit of the model to the empirical data, Figure~\ref{figShrinkage} shows the largest 10 and the remaining 90 shrunken and unshrunken targeted intercepts $\hat\domegai$ from the score-driven transition dynamics in \eqref{eq:first score eq}.
For the largest eigenvalues we observe that shrinkage leads to a (slight) downward adjustment, since the sample eigenvalues have an upward bias.
For the lowest eigenvalues, the converse happens. 
The \textit{relative} impact of shrinkage is much larger at the lower end of the spectrum.
It is precisely this correction at the lower end of the spectrum that causes the substantial increases in out-of-sample log-likelihood.
To see this, Figure~\ref{figShrinkage} plots the improvement in \textit{out-of-sample} log-likelihood for the dynamic skew $t$ copula due to non-linear shrinkage as a function of the spectral index.
We calculate this change in log-likelihood by only replacing the sample eigenvalues with regularized eigenvalues up to spectral index $i$, for $i=1,\ldots,100$, such that $i=100$ corresponds to the fully regularized model.
The figure shows that the bulk of the out-of-sample likelihood improvement comes from the highest spectral indices ($i>70$).
It underlines the importance of the regularization step for in the new model's set-up for the small eigenvalues jointly with introducing dynamics for the largest eigenvalues.

\section{Conclusion}
\label{sec:Concl}

In this article, we proposed a new dynamic copula model for high-dimensional time-dependent financial applications based on the skewed $t$ copula.
The model does not require variables to be grouped in clusters, but rather uses a spectral decomposition of the copula dependence matrix.
Asymptotic biases in the spectrum are avoided by regularization of the correlation matrix, while the dynamics are captured via time-varying volatilities in the principal spectral dimensions, where the number of dynamic components is kept to a minimum via a model selection procedure based on the BIC.
The combination of all three elements results in a parsimonious, yet flexible time-varying dependence model with limited complexity and increased interpretability. 

A simulation study confirmed that the regularized dynamic copula performs well out-of-sample in high-dimensional settings with general dependence structures. 
In an empirical study, we showed that regularized score-driven spectral dynamics can be used to study contractions in the dependence structure of the international financial market during crisis times, which reduces diversification potential and enhances systemic risk. 
We also showed that regularization outperforms cluster assignments in terms of in-sample and out-of-sample performance for our heterogeneous data set containing a broad range of countries and sectors.
In particular, regularization turned out to be most important at the low end of the spectrum, while the introduction of dynamics was most important for the highest eigenvalues.

\bibliography{references} 

\newpage
\appendix
\begin{center}
{\LARGE\bf Supplementary Materials for\\ ``\mytitle''}
\end{center}
\bigskip
\if1\blind {
  \centerline{\large\textit{Koos Gubbels$^{a}$ and Andre Lucas$^{b}$}}
  \bigskip
  \centerline{$^a$: \large\textit{Tilburg University}}
  \centerline{$^b$: \large\textit{Vrije Universiteit Amsterdam and Tinbergen Institute}}
  \bigskip
} \fi
  
\setcounter{page}{1}
\renewcommand{\thepage}{S\arabic{page}}
\bigskip 
\numberwithin{equation}{section} 
\numberwithin{figure}{section} 
\numberwithin{table}{section} 
\numberwithin{theorem}{section}
\renewcommand{\theequation}{\Alph{section}.\arabic{equation}}
\renewcommand{\thefigure}{\Alph{section}.\arabic{figure}}
\renewcommand{\thetable}{\Alph{section}.\arabic{table}}
\renewcommand{\thetheorem}{\Alph{section}.\arabic{theorem}}

\setcounter{tocdepth}{0}
\addtocontents{toc}{\protect\setcounter{tocdepth}{1}} 
\tableofcontents

\clearpage
\section{Proofs}
\label{app:proofs}

\paragraph{Proof of Proposition~\ref{prop:score dynamics equations}}
Using Eqs.~\eqref{eq:GH copula} and~\eqref{eq:GH pdf} and noting that $\dysit = \dysit(\dPITit, \GHnu, \GHgamma\subi) = \GHcdf\subi^{-1}(\dPITit;\vthetact)$ depends on $\GHnu$ and $\GHgamma\subi$, but not on $\vft$, we obtain
\begin{align}
    \nonumber
    \dnablait &=
    \frac{\partial \copulac(\vPITt\mid \calFtm,\vthetact)}{\partial\dfit} 
    =
    \frac{\partial}{\partial\dfit}\log\GHpdf(\vyst;\vthetact)
    - \sum_{\dimi=1}^{\dimN} \frac{\partial}{\partial\dfit}\log\GHpdf\subi(\dysit;\vthetact)
    \\
    \nonumber
    &=
    -\tfrac12
    \frac{\partial\log|\GHmRt|}{\partial\dfit}
    + \vyst{}\trans \frac{\partial\GHvbetat}{\partial\dfit}
    -\frac{(\dimN+\GHnu)}{2}\frac{\partial\log\bra{\GHnu + \vyst{}\trans\GHmRt^{-1}\vyst}}{\partial\dfit}
    + \frac{\partial\log\bra{\GHalphatildet^{(\dimN+\GHnu)/2}\, K_{(\dimN+\GHnu)/2}\bra{\GHalphatildet} }}{\partial\dfit}
    ,
\end{align}
where $\GHalphat = \sqrt{\GHvgamma\trans\GHmRt^{-1}\GHvgamma}$, $\GHvbetat = \GHmRt^{-1}\GHvgamma$, and $\GHalphatildet = \GHalphat\cdot\sqrt{\GHnu + \vyst{}\trans\GHmRt^{-1}\vyst}$.
Defining $k'_\GHnu(x) = \partial\log(x^\GHnu\,K_\GHnu(x))/\partial x$ and $\dwt = (\dimN+\GHnu)/(\GHnu + \vyst{}\trans\GHmRt^{-1}\vyst)$, we obtain
\begin{align}
    \nonumber
    \dnablait &=
    -\tfrac12
    \frac{\partial\log|\GHmRt|}{\partial\dfit}
    + \vyst{}\trans \frac{\partial\GHmRt^{-1}}{\partial\dfit}\GHvgamma
    -\tfrac12\ \frac{\dimN+\GHnu}{\GHnu + \vyst{}\trans\GHmRt^{-1}\vyst}
    \vyst{}\trans\frac{\partial\GHmRt^{-1}}{\partial\dfit}\vyst
    + k'_{(\dimN+\GHnu)/2}\bra{\GHalphatildet}\frac{\partial\GHalphatildet}{\partial\dfit},
    \\
    \nonumber
    &=
    -\tfrac12
    \frac{\partial\log|\GHmRt|}{\partial\dfit}
    + \vyst{}\trans \frac{\partial\GHmRt^{-1}}{\partial\dfit}\GHvgamma
    -\tfrac12\dwt\,
    \vyst{}\trans\frac{\partial\GHmRt^{-1}}{\partial\dfit}\vyst
    + \tfrac12\GHalphatildet\,
    k'_{(\dimN+\GHnu)/2}\bra{\GHalphatildet}\cdot\bra{
        \frac{\GHvgamma\trans\frac{\partial\GHmRt^{-1}}{\partial\dfit}\GHvgamma}{\GHvgamma\trans\GHmRt^{-1}\GHvgamma}
        +
        \frac{\vyst{}\trans\frac{\partial\GHmRt^{-1}}{\partial\dfit}\vyst}{\nu+\vyst{}\trans\GHmRt^{-1}\vyst}
    }
    .
\end{align}
Defining $\mSigmat = \mW\mLambdat\mW\trans$ and $\mDeltaSigmat=\diag(\mSigmat)$ as a diagonal matrix holding the diagonal of $\mSigmat$.
With $\dfit = \log\dlambdait$, the result now follows by noting that
\begin{align}
    \nonumber
    &\GHmRt^{-1} 
    = 
    \mDeltaSigmat^{1/2}\ \mW \mLambdat^{-1} \mW\trans\ \mDeltaSigmat^{1/2}
    ,
    \\
    \nonumber
    &\mW\,\frac{\partial\mLambda^{-1}}{\partial\dfit}\,\mW\trans
    =
    -\,\frac{\vwi\vwi\trans}{\dlambdait}
    ,
    \\
    \nonumber
    &\frac{\partial\dSigmajjt}{\partial\dfit}
    =
    \frac{\partial}{\partial\dfit}
    \sum_{\dimk=1}^{\dimN} \dlambda_{\dimk,\dimt} \dW_{\dimj,\dimk}^2
    =
    \dlambdait \dWji^2
    ,
    \\
    \nonumber
    &\frac{\partial \log|\GHmRt|}{\partial\dfit} 
    = 
    \frac{\partial \log|\mLambdat|}{\partial\dfit} 
    -
    \frac{\partial \log|\mDeltaSigmat|}{\partial\dfit}
    = 
    1 - \sum_{\dimj=1}^{\dimN} \frac1{\dSigmajjt}\,\frac{\partial\dSigmajjt}{\partial\dfit}
    = 
    1 - \dlambdait \sum_{\dimj=1}^{\dimN} \frac{\dWji^2}{\dSigmajjt}
    ,
    \\
    \nonumber
    &\frac{\partial \GHmRt^{-1}}{\partial\dfit}
    = -\mDeltaSigmat^{1/2}\,\frac{\vwi\vwi\trans}{\dlambdait}\,\mDeltaSigmat^{1/2}
    + \mSigmadotit\,\GHmRt^{-1} + \GHmRt^{-1}\,\mSigmadotit\trans
    ,
    \\
    \nonumber
    &\mSigmadotit
    = 
    \mSigmadotit\trans
    =
    \frac{\partial\mDeltaSigmat^{1/2}}{\partial\dfit}\,
    \mDeltaSigmat^{-1/2}
    =
    \tfrac12\,\dlambdait\,\diag\bra{
        \frac{\dW_{1,\dimi}^2}{\dSigma_{1,1,\dimt}},
        \ldots,
        \frac{\dW_{\dimN,\dimi}^2}{\dSigma_{\dimN,\dimN,\dimt}},
    }
    .
\end{align}
Since $\partial K_\GHnu(x)/\partial x = -1/2(K_{\GHnu+1}(x)+K_{\GHnu-1}(x))$, all derivatives are determined analytically.
This proves the result. \hfill\qedsymbol

\bigskip

\paragraph{Proof of Eq.~\eqref{eq:GH score expression rewrite}}
Define $\vei$ as the $\dimi$th column from the identity matrix, and $\mDeltawi$ as a diagonal matrix holding the $\dimi$th eigenvector $\vwi$ from $\mW$.
Using the definitions
\begin{align*}
    \vystildet &= \mW\trans\mDeltaSigmat^{1/2}\vyst,
    \qquad
    \GHvgammatildet = \mW\trans\mDeltaSigmat^{1/2}\GHvgamma,
    \qquad
    \vysbarit = \mW\trans\mDeltaSigmat^{1/2}\mSigmadotit\vyst,
    \qquad
    \GHvgammabarit = \mW\trans\mDeltaSigmat^{1/2}\mSigmadotit\GHvgamma,
    \\
    \GHalphat &= 
    \sqrt{\GHvgamma\trans\GHmRt^{-1}\GHvgamma}=
    \sqrt{\GHvgamma\trans\mDeltaSigmat^{1/2}\mW\mLambdat^{-1}\mW\trans\mDeltaSigmat^{1/2}\GHvgamma}=
    \sqrt{\GHvgammatildet\trans \mLambdat^{-1} \GHvgammatildet},
    \\
    \GHalphatildet &= 
    \GHalphat\cdot\sqrt{\GHnu + \vyst{}\trans\GHmRt^{-1}\vyst}
    =
    \GHalphat\cdot\sqrt{\GHnu + \vyst{}\trans\mDeltaSigmat^{1/2}\mW\mLambdat^{-1}\mW\trans\mDeltaSigmat^{1/2}\vyst}
    =
    \GHalphat\cdot\sqrt{\GHnu + \vystildet{}\trans\mLambdat^{-1}\vystildet}
    ,
    \\
    \dwt &=
    (\GHnu+\dimN)/(\GHnu + \vyst{}\trans\GHmRt^{-1}\vyst)
    =
    (\GHnu+\dimN)/(\GHnu + \vystildet{}\trans\mLambdat^{-1}\vystildet)
    ,
    \\
    \mSigmatildeit &= \mSigmadotit\GHmRt^{-1} + \GHmRt^{-1}\mSigmadotit
    , 
    \qquad
    \mSigmadotit = \tfrac12\,\dlambdait\, \mDeltaSigmat^{-1}\mDeltawi^2,
    \\
    \frac{\partial\GHmRt^{-1}}{\partial\dfit} &=  
    - \dlambdait^{-1}\,
    \mDeltaSigmat^{1/2} \mW\vei \vei\trans \mW\trans \mDeltaSigmat^{1/2}
    + \mSigmatildeit
    ,
\end{align*}
we obtain
\begin{align*}
    \GHvgamma\trans\mSigmatildeit\GHvgamma 
    &=
    \GHvgamma\trans\mSigmadotit\mDeltaSigmat^{1/2}\mW\ \mLambdat^{-1}\ \mW\trans\mDeltaSigmat^{1/2}\GHvgamma 
    + 
    \GHvgamma\trans\mDeltaSigmat^{1/2}\mW\ \mLambdat^{-1}\ \mW\trans\mDeltaSigmat^{1/2}\mSigmadotit\GHvgamma
    \\&= 
    \GHvgammabarit\trans\mLambdat^{-1}\GHvgammatildet + \GHvgammatildet\trans\mLambdat^{-1}\GHvgammabarit
    =
    2\GHvgammatildet\trans\mLambdat^{-1}\GHvgammabarit
    ,
    \\\\
    \GHvgamma\trans\mSigmatildeit\vyst 
    &=
    \GHvgamma\trans\mSigmadotit\mDeltaSigmat^{1/2}\mW\ \mLambdat^{-1}\ \mW\trans\mDeltaSigmat^{1/2}\vyst 
    + 
    \GHvgamma\trans\mDeltaSigmat^{1/2}\mW\ \mLambdat^{-1}\ \mW\trans\mDeltaSigmat^{1/2}\mSigmadotit\vyst 
    \\&= 
    \GHvgammabarit\trans\mLambdat^{-1}\vystildet + \GHvgammatildet\trans\mLambdat^{-1}\vysbarit
    ,
    \\\\
    \vyst{}\trans\mSigmatildeit\vyst 
    &=
    \vyst{}\trans\mSigmadotit\mDeltaSigmat^{1/2}\mW\ \mLambdat^{-1}\ \mW\trans\mDeltaSigmat^{1/2}\vyst 
    + 
    \vyst{}\trans\mDeltaSigmat^{1/2}\mW\ \mLambdat^{-1}\ \mW\trans\mDeltaSigmat^{1/2}\mSigmadotit\vyst 
    \\&= 
    \vysbarit{}\trans\mLambdat^{-1}\vystildet + \vystildet{}\trans\mLambdat^{-1}\vysbarit
    = 
    2 \vysbarit{}\trans\mLambdat^{-1}\vystildet
    ,
    \\\\
    \GHvgamma\trans \frac{\partial\GHmRt^{-1}}{\partial\dfit}\GHvgamma
    &= 
    -\dlambdait^{-1}\,\GHvgammatildet\trans\vei\vei\trans\GHvgammatildet
    +
    \GHvgamma\trans\mSigmatildeit\GHvgamma
    = 
    -\,\frac{\GHgammatildeit^2}{\dlambdait}
    +
    2\GHvgammatildet\trans\mLambdat^{-1}\GHvgammabarit
    ,
    \\\\
    \GHvgamma\trans \frac{\partial\GHmRt^{-1}}{\partial\dfit}\vyst
    &= 
    -\dlambdait^{-1}\,\GHvgammatildet\trans\vei\vei\trans\vystildet
    +
    \GHvgamma\trans\mSigmatildeit\vyst
    = 
    -\,\frac{\GHgammatildeit\dystildeit}{\dlambdait}
    +
    \GHvgammabarit\trans\mLambdat^{-1}\vystildet
    +
    \GHvgammatildet\trans\mLambdat^{-1}\vysbarit
    ,
    \\\\
    \vyst{}\trans \frac{\partial\GHmRt^{-1}}{\partial\dfit}\vyst
    &= 
    -\dlambdait^{-1}\,\vystildet{}\trans\vei\vei\trans\vystildet
    +
    \vyst{}\trans\mSigmatildeit\vyst
    = 
    -\,\frac{\dystildeit{}^2}{\dlambdait}
    +
    2\vystildet{}\trans\mLambdat^{-1}\vysbarit
    .
\end{align*}
We can now rewrite \eqref{eq:GH score expression} as
\begin{align*}
    \label{eq:GH score expression rewrite}
        \dnablait &=
        \bra{
            -\tfrac12 + 
            \tfrac12\dlambdait \sum_{\dimj=1}^{\dimN}
            \frac{\dWji^2}{\dSigmajjt}
        }
        +
        \bra{
            -\,\frac{\GHgammatildeit\dystildeit}{\dlambdait}
            +\GHvgamma\trans\mSigmatildeit\vyst
        }
        +
        \bra{
            \tfrac12\,\dwt\,\frac{\dystildeit{}^2}{\dlambdait}
            -\tfrac12\,\dwt\,\vyst{}\trans\mSigmatildeit\vyst
        }
        \\ & \qquad
            -\tfrac12\,
            \GHalphatildet\cdot
            k'_{(\GHnu+\dimN)/2}\bra{\GHalphatildet} \cdot
            \bra{
                \frac{\GHgammatildeit^2/\dlambdait}{\GHvgammatildet\trans\mLambdat^{-1}\GHvgammatildet}
                +
                \frac{\dystildeit{}^2/\dlambdait}{\nu+\vystildet{}\trans\mLambdat^{-1}\vystildet}
            }
            \\ &\qquad
            +\tfrac12\,
            \GHalphatildet\cdot
            k'_{(\GHnu+\dimN)/2}\bra{\GHalphatildet} \cdot
            \bra{
                \frac{\GHvgamma\trans\mSigmatildeit\GHvgamma}{\GHvgammatildet\trans\mLambdat^{-1}\GHvgammatildet}
                +
                \frac{\vyst{}\trans\mSigmatildeit\vyst}{\nu+\vystildet{}\trans\mLambdat^{-1}\vystildet}
            }
        \\\\
        &=
        \tfrac12\,\bra{
            \dwt\,\frac{\dystildeit{}^2}{\dlambdait}
            - 1 - 2\frac{\GHgammatildeit\dystildeit}{\dlambdait}
        }
        -\tfrac12\,
        \GHalphatildet\cdot
        k'_{(\GHnu+\dimN)/2}\bra{\GHalphatildet} \cdot
        \bra{
            \frac{\GHgammatildeit^2/\dlambdait}{\GHvgammatildet\trans\mLambdat^{-1}\GHvgammatildet}
            +
            \frac{\dystildeit{}^2/\dlambdait}{\nu+\vystildet{}\trans\mLambdat^{-1}\vystildet}
        }
        \\ &\qquad
        -\,\bra{
            \dwt\,\vystildet{}\trans\mLambdat^{-1}\vysbarit
            - 
            \trace(\mSigmadotit)
            - \GHvgammatildet{}\trans\mLambdat^{-1}\vysbarit
            - \GHvgammabarit\trans\mLambdat^{-1}\vystildet
        }
        \\ &\qquad
        +
        \GHalphatildet\cdot
        k'_{(\GHnu+\dimN)/2}\bra{\GHalphatildet} \cdot
        \bra{
            \frac{\GHvgammatildet\trans\mLambdat^{-1}\GHvgammabarit}{\GHvgammatildet\trans\mLambdat^{-1}\GHvgammatildet}
            +
            \frac{\vystildet{}\trans\mLambdat^{-1}\vysbarit}{\nu+\vystildet{}\trans\mLambdat^{-1}\vystildet}
        }.
\end{align*}
This proves the result.\hfill\qedsymbol

\bigskip

\clearpage
\section{Supplementary material simulation study}
\label{app:additional simulations}

\subsection{Reference model}
\label{app:oh methodology}
As a reference model in our simulation and empirical studies, we use the dynamic factor copula model, which was introduced by \cite{Opschoor2020} and refined by \cite{Oh2023}. The factor loadings are normalized as:
\begin{equation}
\tilde{\bm{\lambda}}_{i,t} = \frac{\bm{\lambda}_{i,t}}{\sqrt{1+ \bm{\lambda}'_{i,t}\bm{\lambda}_{i,t}}}, \quad \sigma^2_{i,t}=\frac{\bm{\lambda}_{i,t}}{\sqrt{1+ \bm{\lambda}'_{i,t}\bm{\lambda}_{i,t}}}
\end{equation}
The copula correlation matrix is given by:
\begin{equation}
\bm{R}_t = \tilde{\bm{L}}_{t}\tilde{\bm{L}}_{t}+\bm{D}_t
\end{equation}
Here, we have that  $\tilde{\bm{L}}_{t}=[\tilde{\bm{\lambda}}_{1,t},...,\tilde{\bm{\lambda}}_{d,t}]$ and $\bm{D}_t= {\rm diag}(\sigma^2_{1,t},...,\sigma^2_{d,t})$.

In particular, we use the following multi-factor structure considered by \cite{Opschoor2020} and \cite{Oh2023} with $2n_G$ factor loadings.
Suppose there are $n_G = 25$ groups with 4 assets per group. The matrix $\tilde{\bm{L}}_{t}$ is then generated by the following tensor product of a matrix with a vector: 
\begin{equation}\label{eqCluster}
\tilde{\bm{L}}_{t}=\begin{pmatrix}
\tilde{\lambda}^M_{1,t} & \tilde{\lambda}^C_{1,t} & 0 &...  & 0 \\ 
\tilde{\lambda}^M_{2,t} & 0 & \tilde{\lambda}^C_{2,t} & ... & 0\\   
... & ... & ... & ... & ...\\
\tilde{\lambda}^M_{n_G,t} & 0 & 0 & ... & \tilde{\lambda}^C_{n_G,t} \\
\end{pmatrix} 
\otimes
\begin{pmatrix}
1  \\ 
1  \\   
1  \\   
1
\end{pmatrix}
\end{equation}

\cite{Oh2023} have developed a method to optimally group stocks into clusters. We use their algorithm to determine the optimal number of clusters and the optimal assignment of indices.

The dynamics of the reference model is given by \citep{Oh2023}
\begin{eqnarray}
\lambda^M_{g,t+1}&=&\omega^M_g+\alpha^M\frac{\partial \log c(\bm{u_t};\bm{R}_t,\nu,\bm{\gamma})}{\partial \lambda^M_{g,t}}+\beta^M\lambda^M_{g,t},\nonumber\\
\lambda^C_{g,t+1}&=&\omega^C_g+\alpha^C\frac{\partial \log c(\bm{u_t};\bm{R}_t,\nu,\bm{\gamma})}{\partial \lambda^C_{g,t}}+\beta^C\lambda^C_{g,t}.
\end{eqnarray}
We estimate the model by \cite{Oh2023} on our data set using their code. The parameter estimation results are shown in Table \ref{tabEstClus100D}.

\clearpage
\subsection{Experiment: mis-specification}
\label{app:sim misspec}

\begin{figure}[h]
\centering
\includegraphics[width=0.7\columnwidth]{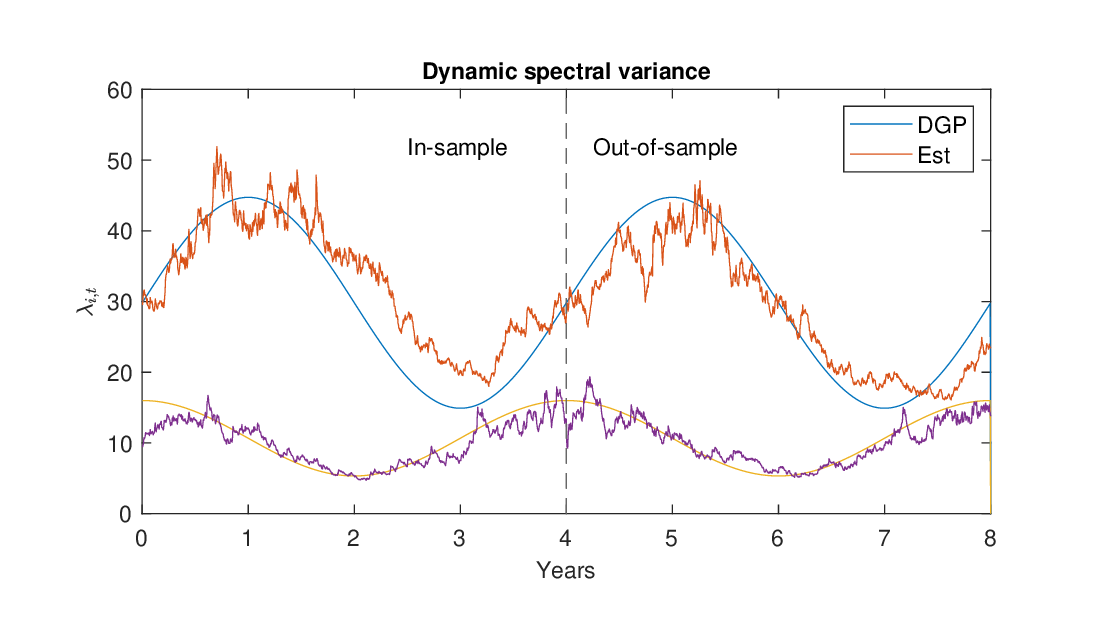}
\caption{Dynamics of first and second eigenvalue, where the true dynamics of the dgp are based on a prespecified periodic pattern, while the estimated model is based on GAS. The true dynamics are compared to the predicted dynamics from the estimated model. The period after 4 years is an out-of-sample forecast. }\label{sim:figEigValDynMis}
\end{figure}

We perform an additional simulation experiment to show that the skew $t$ copula with score-driven eigenvalues can recover general patterns over time, even if the model is mis-specified; see Figure~\ref{sim:figEigValDynMis}.
In the experiment, we used the same model parameters as in Section \ref{subsec:SimDynamic}, but now the true dynamics are given by periodic patterns over time: $\lambda_{1,t}=\lambda_1(1+\sin(4\pi t/T)/2)$ and $\lambda_{2,t}=\lambda_2(1+\cos(4\pi t/T)/2)$. 
We estimate the model using regularized spectral score-driven dynamics with Eq.~(\ref{eq:loglik iteration k}).
We find that the model recovers the true dynamics of the eigenvalues well, even though the model was not informed about the periodic specification.

\clearpage
\section{Supplementary material empirical study}
\label{app:additional empirics}

In Table \ref{tabIndices100}, we list 100 stocks based on 10 countries and 10 sectors.
These stocks are used in the empirical analyses.
In a few cases, there was no stock available for the country-sector combination based on the requirement of a full data history.
Consequently, other country-sector combinations have been selected twice.
In total, each country and each sector occurs precisely 10 times in the data set.

\begin{table}[h]
\caption{\label{tabIndices100} RIC codes of 100 stocks that are included in the empirical study. The first column shows the two-letter country code, while the first row and the middle row show the industry sector.  }
\begin{center}
\begin{tabular}{l || l l l l l} 
 & Financials & Industrials &  Basic  & Consumer & Utilities  \\ 
 &  &  &   Materials & Non- Cyclicals & \\
\hline
\hline
DE & ALVG &  SIEG & BASF & BEIG & EONG   \\ 
FR & AXAF &  SCHN & AIRP & OREP & ENGIE  \\  
UK & HSBA &  RR   & RIO  & ULVR & NG     \\
ES & SAN  &  ACS  & VID  & EBRO & IBE,\, ELE    \\ 
IT & CRDI &  LDOF & BZU  & CPRI & ENEI,\, TRN   \\
SW & INVE &  ATCO & SAND & LIFCO &    \\ 
NO & DNB  &  KOG  & NHY  & ORK  & SCATC   \\
NL & INGA &  WLSN & AKZO & HEIN &    \\  
BE & KBC  &  ACKB & UMI  & ABI  & ELI   \\ 
CH & UBSG &  ABBN & HOLN & NESN & BKWB   \\
\hline
\hline
 &  Consumer & Technology & Health Care & Energy & Real Estate \\ 
 &  Cyclicals &  & &  &  \\
\hline
\hline
DE & BMWG &  SAPG & MRCG & PNEG & VNA  \\ 
FR & LVMH &  ORAN & ESLX & TTEF & LOIM \\  
UK & CPG  &  RELNG & AZN  & SHEL & SGRO \\
ES & ITX  &  AMA  &    & REP & MRL \\
IT & MONC &  TLIT & RECI & ENI &  \\ 
SW & ASSA,\, HM &  HEXA & SOBIV,\, SECT & & SAGA \\ 
NO & VENDA &  TEL &  & EQNR,\, AKRBP & OLT \\
NL &  & ASML & PHG & VOPA,\, SBMO & ECMPA,\, WEHA \\  
BE & IETB &  MLXS & UCB & CMBT & WDPP \\ 
CH & CFR &  SCMN & NOVN,\, RO &  & SPSN \\
\hline
\end{tabular}
\end{center}
\end{table}

\clearpage
Table \ref{tabEstClus100D} shows the estimated parameters for the static skew $t$ copula with estimated cluster assignments for empirical study in 100 dimensions.

\begin{table} [h] \centering
\caption{\label{tabEstClus100D} Parameter estimation of the static skew $t$ copula for empirical study with 100 stocks using optimal cluster assignments.}
\begin{tabular}{l c c l c c l c c l c} 
\hline
Par     & Est &  & Par     & Est &  & Par & Est &  & Par & Est  \\
\midrule
$\omega^M_1$  & 0.99 &  & $\omega^M_{10}$  & 1.23 &  &$\omega^C_1$ & 0.002 &  & $\omega^C_{10}$  & 0.26    \\
$\omega^M_2$  & 1.26 &  & $\omega^M_{11}$  & 0.29 &  &$\omega^C_2$ & 0.53  &  & $\omega^C_{11}$  & 0.20   \\
$\omega^M_3$  & 0.721 &  & $\omega^M_{12}$  & 0.87 &  &$\omega^C_3$ & 0.68 &  & $\omega^C_{12}$  & 0.78    \\
$\omega^M_4$  & 0.84 &  & $\omega^M_{13}$  & 0.65 &  &$\omega^C_4$ & 0.26 &  & $\omega^C_{13}$  & 0.32   \\
$\omega^M_5$  & 0.52 &  & $\omega^M_{14}$  & 0.88 &  &$\omega^C_5$ & 0.21 &  & $\omega^C_{14}$  & 0.90   \\
$\omega^M_6$  & 0.67 &  & $\omega^M_{15}$  & 0.72 &  &$\omega^C_6$ & 0.004 &  & $\omega^C_{15}$  & 0.47   \\ 
$\omega^M_7$  & 1.21 &  & $\omega^M_{16}$  & 0.41 &  &$\omega^C_7$ & 0.87 &  & $\omega^C_{16}$  & 0.28    \\
$\omega^M_8$  & 0.73 &  & $\omega^M_{17}$  & 0.77 &  &$\omega^C_8$ & 0.52 &  & $\omega^C_{17}$  & 0.52   \\ 
$\omega^M_9$  & 0.80 &  &   $\nu$          & 33.9 &  &$\omega^C_9$ & 0.59 &  & $\gamma$  &  -0.23  \\ 
\hline
\end{tabular}
\end{table}

\bigskip

Finally, Figure \ref{fig:omega OhPatton vs eigvec} compares the loadings of the cluster-based factor copula of \citet{Oh2023} versus the elements of the first and second spectral eigenvector.

\bigskip

\begin{figure}[h]\centering
  \begin{subfigure}[t]{0.48\textwidth}
  \centering
    \includegraphics[width=\linewidth]{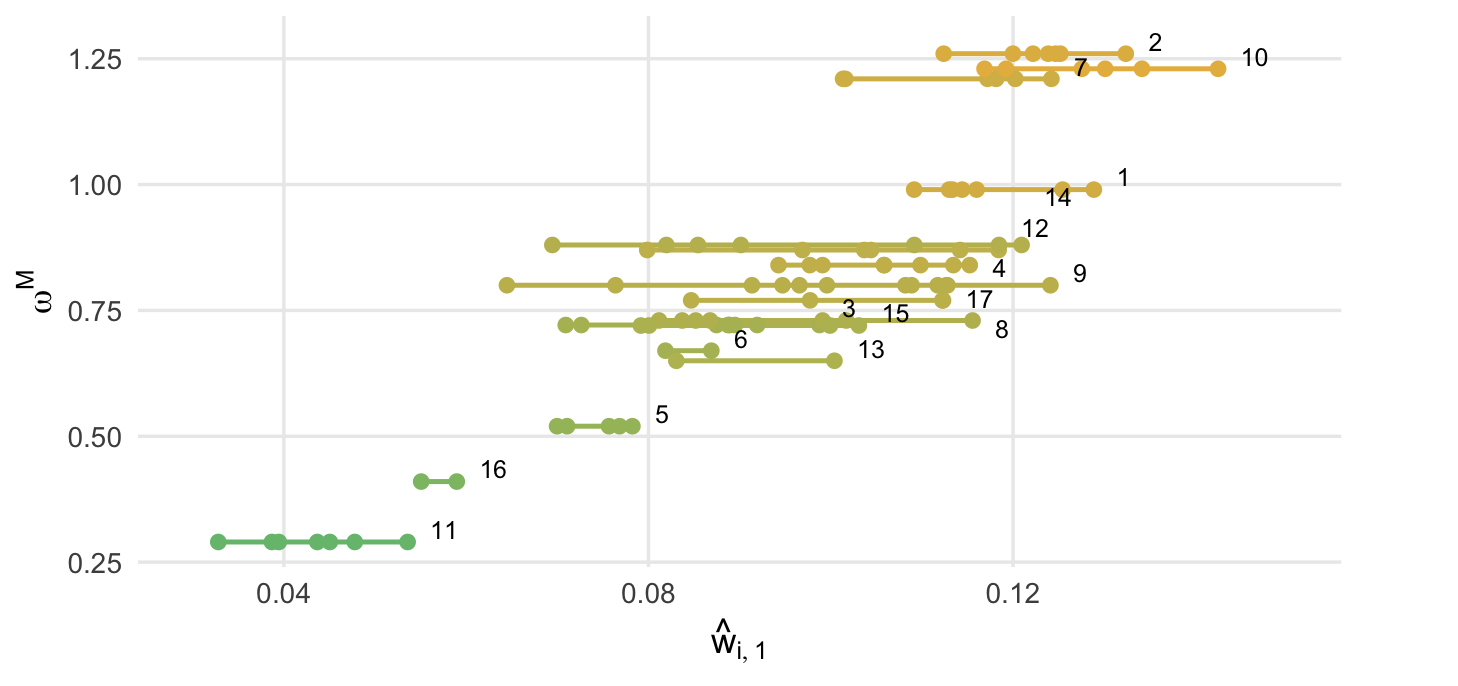}
    \caption{$\omega_M$ versus first spectral eigenvector $\hat w_{i,1}$}
    \label{fig:omega OhPatton vs eigvec 1}
  \end{subfigure}
  \hfill
  \begin{subfigure}[t]{0.48\textwidth}
    \centering
    \includegraphics[width=\linewidth]{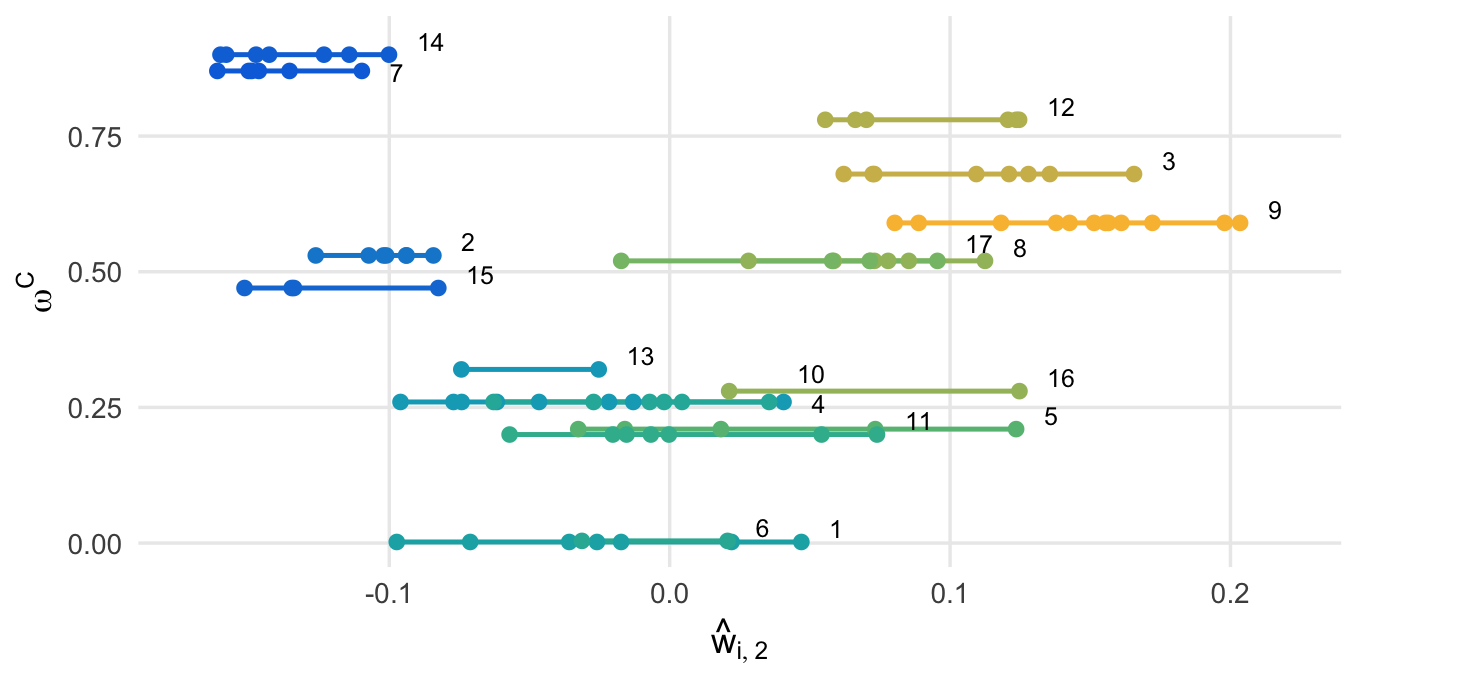}
    \caption{$\omega_C$ versus first spectral eigenvector $\hat w_{i,2}$}
    \label{fig:omega OhPatton vs eigvec 2}
  \end{subfigure}
  \label{fig:omega OhPatton vs eigvec}
\caption{Loadings of the cluster-based factor copula of \citet{Oh2023} versus the elements of the first and second spectral eigenvectors. Elements are grouped by cluster number 1\ldots17, which is the BIC optimal number of clusters. Each dot corresponds to an single stock, and the colors correspond to the average value of the loadings as in Figure~\ref{figEigVec}.
The loadings of the first eigenvector of the spectral copula and the loadings of the first factor in the factor copula almost linearly relate.
The magnitudes of the second eigenvector elements of the spectral approach and those of the individual clusters of the factor copula group factors, also clearly relate, where differences in signs can be captured by the signs of the different group factors in the factor copula approach. Note that all second eigenvector elements in the spectral approach are weighted by $\dmu_{2,\dimt}$, such that sign differences must be reflected in the eigenvector elements themselves.
\label{fig:omega OhPatton vs eigvec}}
\end{figure}

\end{document}

%% file: notation.tex
\newcommand{\mymbf}[1]{\bm{#1}}

\def\dimi{i}
\def\dimj{j}
\def\dimk{k}
\def\dimN{d}
\def\dimNz{\dimN_0}
\def\dimt{t}
\def\dimT{T}

\def\SR{\mathbb{R}}

\def\Ee{\mathbb{E}}

\def\LL{\mathcal{L}}

\def\staticpar{\psi}
\def\vstaticpar{\mymbf{\staticpar}}

\def\subi{_{\dimi}}

\def\subij{_{\dimi,\dimj}}
\def\subji{_{\dimj,\dimi}}
\def\subj{_{\dimj}}
\def\subt{_{\dimt}}
\def\subtm{_{\dimt-1}}
\def\subtp{_{\dimt+1}}
\def\subit{_{\dimi,\dimt}}
\def\subitp{_{\dimi,\dimt+1}}
\def\subjt{_{\dimj,\dimt}}

\def\subjjt{_{\dimj,\dimj,\dimt}}

\def\copulac{c}

\def\PIT{u}
\def\dPITit{\PIT\subit}
\def\vPIT{\mymbf{\PIT}}
\def\vPITt{\vPIT\subt}

\def\cdfy{F}
\def\cdfyi{\cdfy\subi}

\def\GHpdf{g}

\def\GHcdf{G}
\def\GHnu{\nu}
\def\GHalpha{\alpha}
\def\GHalphat{\GHalpha\subt}
\def\GHalphatilde{\tilde\GHalpha}
\def\GHalphatildet{\GHalphatilde\subt}
\def\GHbeta{\beta}

\def\GHvbeta{\mymbf{\GHbeta}}
\def\GHvbetat{\GHvbeta\subt}

\def\GHgamma{\gamma}
\def\GHgammai{\GHgamma\subi}
\def\GHgammatilde{\tilde{\GHgamma}}

\def\GHgammatildeit{\GHgammatilde\subit}
\def\GHvgamma{\mymbf{\GHgamma}}
\def\GHvgammatilde{\tilde{\GHvgamma}}
\def\GHvgammatildet{\GHvgammatilde\subt}
\def\GHvgammabar{\bar{\GHvgamma}}
\def\GHvgammabarit{\GHvgammabar\subit}
\def\GHdR{R}

\def\GHmR{\mymbf{\GHdR}}
\def\GHmRt{\GHmR\subt}

\def\GHmRshrink{\check{\GHmR}}
\def\GHmRshrinkt{\GHmRshrink\subt}

\def\da{a}
\def\dai{\da\subi}
\def\mA{\mymbf{A}}

\def\db{b}
\def\dbi{\db\subi}
\def\mB{\mymbf{B}}

\def\ve{\mymbf{e}}
\def\vei{\ve\subi}

\def\df{f}
\def\dfit{\df\subit}

\def\vf{\mymbf{\df}}
\def\vft{\vf\subt}
\def\vftp{\vf\subtp}

\def\calF{\mathcal{F}}
\def\calFt{\calF\subt}
\def\calFtm{\calF\subtm}

\def\dh{h}

\def\Nn{{\rm N}}

\def\dq{q}

\def\dv{\varv}
\def\dvt{\dv\subt}

\def\dw{\tilde\nu}
\def\dwt{\dw\subt}
\def\dW{w}

\def\dWji{\dW\subji}
\def\vw{\mymbf{w}}
\def\vwi{\vw\subi}
\def\mW{\mymbf{W}}

\def\dy{y}
\def\dyit{\dy\subit}
\def\vy{\mymbf{\dy}}
\def\vyt{\vy\subt}

\def\Covy{\mymbf{S}}
\def\Covys{\Covy^{\star}}
\def\Covyshat{\hat{\Covy}^{\star}}

\def\dys{y^{\star}}
\def\vystilde{\tilde{\mymbf{y}}^{\star}}
\def\vysbar{\bar{\mymbf{y}}^{\star}}
\def\dysit{\dys\subit}
\def\vys{\mymbf{\dys}}
\def\vyst{\vys\subt}
\def\dystilde{\tilde y^{\star}}
\def\dystildeit{\dystilde\subit}
\def\vystildet{\vystilde\subt}
\def\vysbarit{\vysbar\subit}

\def\vz{\mymbf{z}}
\def\vzt{\vz\subt}

\def\dnabla{\nabla}
\def\dnablait{\dnabla\subit}
\def\vnabla{\mymbf{\dnabla}}
\def\vnablat{\vnabla\subt}

\def\SIMbeta{\beta}
\def\SIMmarket{M}
\def\SIMgroup{G}
\def\SIMcountry{C}
\def\SIMidio{I}
\def\SIMgroupi{\SIMgroup\subi}
\def\SIMgroupj{\SIMgroup\subj}
\def\SIMcountryi{\SIMcountry\subi}
\def\SIMcountryj{\SIMcountry\subj}
\def\SIMbetaM{\SIMbeta_{\SIMmarket}}

\def\SIMbetaGi{\SIMbeta_{\SIMgroupi}}

\def\SIMbetaGj{\SIMbeta_{\SIMgroupj}}
\def\SIMbetaC{\SIMbeta_{\SIMcountry}}
\def\SIMbetaCi{\SIMbeta_{\SIMcountryi}}

\def\SIMbetaI{\SIMbeta_{\SIMidio}}

\def\krondelta{\delta}

\def\dlambda{\lambda}
\def\dlambdait{\dlambda\subit}

\def\dlambdahat{\hat\dlambda}

\def\vlambda{\mymbf{\dlambda}}
\def\vlambdat{\vlambda\subt}
\def\vlambdahat{\hat\vlambda}
\def\vlambdahatt{\vlambdahat\subt}
\def\mLambda{\mymbf{\Lambda}}
\def\mLambdat{\mLambda\subt}

\def\mLambdashrink{\check{\mymbf{\Lambda}}}

\def\dmu{\check{\lambda}}

\def\vmu{\check{\vlambda}}

\def\dtheta{\theta}
\def\vtheta{\mymbf{\dtheta}}

\def\vthetact{\vtheta_{c,\dimt}}

\def\viota{\mymbf{\iota}}

\def\dSigma{\Sigma}

\def\dSigmajjt{\dSigma\subjjt}
\def\mSigma{\mymbf{\Sigma}}
\def\mSigmat{\mSigma\subt}

\def\mSigmatildeit{\tilde\mSigma\subit}
\def\mSigmadot{\dot\mSigma}
\def\mSigmadotit{\mSigmadot\subit}\def\mSigmashrink{\check{\mSigma}}
\def\mSigmashrinkt{\mSigmashrink\subt}

\def\mDelta{\mymbf{\Delta}}

\def\mDeltaSigmat{\mDelta_{\mSigmat}}
\def\mDeltawi{\mDelta_{\vwi}}

\def\domega{\omega}
\def\domegai{\omega\subi}
\def\vomega{\mymbf{\domega}}

\def\factorsymbol{F}
\def\factorMt{\factorsymbol^{\SIMmarket}\subt}
\def\factorGit{\factorsymbol^{\SIMgroupi}\subt}
\def\factorCit{\factorsymbol^{\SIMcountryi}\subt}
\def\factorepsit{\eta\subit}

\newcommand{\bra}[1]{\left( #1 \right)}
\newcommand{\sbra}[1]{\left[ #1 \right]}

\def\trans{^\top}
\def\cfdot{\,\cdot\,}

\def\unitmat{\mathbf{I}}
\def\unitmatN{\unitmat_{\dimN}}
\def\vzero{\mymbf{0}}
\def\vzeroN{\vzero_{\dimN}}

\DeclareMathOperator{\diag}{diag}
\DeclareMathOperator{\trace}{trace}

\def\qedsymbol{\rule{1ex}{1ex}}